\documentclass[twocolumn,amsmath,amssymb,floatfix,superscriptaddress,prl]{revtex4-2}

\usepackage{bm,graphicx,url,epsf,color}
\usepackage{amssymb}
\usepackage{amsthm}
\usepackage{xcolor}
\usepackage{comment}
\usepackage{relsize}
\usepackage{xspace}
\usepackage[normalem]{ulem}
\usepackage[super]{nth}
\usepackage{braket}
\usepackage{soul}
\usepackage{textcomp}
\usepackage[T1]{fontenc}
\usepackage{stmaryrd}
\usepackage{wrapfig}
\usepackage[ruled,lined]{algorithm2e}

\graphicspath{{graphics/}}
\usepackage[colorlinks=true,linktocpage,citecolor=blue,linkcolor=blue]{hyperref}

\newcommand{\bd}[1]{{\boldsymbol{#1}}}

\def\hpsi{\hat{\psi}}
\def\hS{\hat{S}}
\def\hs{\hat{s}}
\def\hH{\hat{H}}
\def\hvartheta{\hat{\vartheta}}
\def\hphi{\hat{\phi}}
\def\hpsi{\hat{\psi}}
\def\hvartheta{\hat{\vartheta}}
\def\hU{\hat{U}}

\let\origaddcontentsline\addcontentsline

\begin{document}
	
	\title{Monte Carlo Solution of the Kondo Model}
	
	\author{Nicolas Paris}
	\thanks{These authors contributed equally to this work.}
	\affiliation{Sorbonne Universit\'e, CNRS, Laboratoire de Physique Th\'eorique de la Mati\`ere Condens\'ee, LPTMC, F-75005 Paris, France}
	\affiliation{Universit\'e Paris Cit\'e, CNRS,  Laboratoire  Mat\'eriaux  et  Ph\'enom\`enes  Quantiques, 75013  Paris,  France}
	
	\author{Oscar Bouverot-Dupuis}
	\thanks{These authors contributed equally to this work.}
	\affiliation{Universit\'{e} Paris Saclay, CNRS, LPTMS, 91405, Orsay, France}
	\affiliation{IPhT, CNRS, CEA, Universit\'{e} Paris Saclay, 91191 Gif-sur-Yvette, France}
	
	\author{Christophe Mora}
	\affiliation{Universit\'e Paris Cit\'e, CNRS,  Laboratoire  Mat\'eriaux  et  Ph\'enom\`enes  Quantiques, 75013  Paris,  France}
	
	\graphicspath{{./figures/}}

	\begin{abstract}
		The Kondo model is a paradigmatic quantum impurity problem realized in a wide variety of experimental platforms and central to the study of strongly correlated electrons.
		We introduce a discrete model that exactly reproduces the multichannel Kondo model and demonstrate that it can be simulated efficiently. Using cluster Monte Carlo algorithms, we completely eliminate critical slowing down, 
		providing direct access to universal crossover functions and transport properties across a broad range of parameters. Remarkably, the same model captures both the weak- and strong-coupling regimes, unifying descriptions traditionally derived in complementary limits and revealing their common origin.
		Our method naturally accommodates large channel numbers, anisotropy, interacting one-dimensional leads, and channel asymmetry, yielding predictions for transport properties in charge-Kondo devices.
	\end{abstract}
	
	\maketitle

	\paragraph{Introduction ---}
	
	Quantum impurity models provide a paradigmatic setting to investigate the effects of strong electronic correlations. They illustrate how the coupling of a few localized degrees of freedom to an extended environment can genuinely change the low-energy properties of the entire system. Among them, the Kondo model and its multichannel generalizations play a central role~\cite{hewson1993kondo,nozieres1974fermi}. Originally introduced to describe a magnetic impurity interacting with one or several electronic reservoirs, Kondo models host a wealth of collective phenomena, including quantum criticality and the emergence of exotic quasiparticles with non-Abelian statistics~\cite{han2022,lopes2020anyons,komijani2020isolating,lotem2022manipulating,gabay2022,gaines2026,komijani2026}. As some of the simplest interacting many-body systems exhibiting nontrivial infrared behavior, they have become a cornerstone of modern condensed-matter physics.

	The Kondo effect has been observed in a wide variety of experimental platforms, ranging from magnetic impurities in metals to semiconductor quantum dots~\cite{goldhaber1998kondo,goldhaber1998kondoPRL,cronenwett1998tunable,nygaard2000kondo,sasaki2004enhanced,ji2000phase,van2000kondo,sasaki2000kondo,simmel1999anomalous,liang2002kondo,park2002coulomb,jeong2001kondo,craig2004tunable,potok2007observation,chorley2012tunable,keller2015universal} and molecular devices~\cite{madhavan1998tunneling,li1998kondo,zhao2005controlling,zhang2013temperature,bork2011tunable,trishin2023tuning,bagchi2024probing}. This remarkable ubiquity is a consequence of the universal nature of Kondo physics: systems with very different microscopic realizations flow to the same low-energy renormalization-group (RG) fixed points and display identical scaling properties. Yet, these different realizations have traditionally been described by complementary Kondo models adapted to distinct parameter regimes. The conventional Kondo model naturally describes the weak-tunneling regime, while charge-Kondo devices provide access to both weak- and strong-tunneling limits~\cite{Matveev_1991,Matveev_1995,furusaki_1995}. Despite major advances from techniques such as the Bethe Ansatz~\cite{andrei1984,tsvelick1985exact,kattel2026}, numerical RG~\cite{wilson_1975,bulla2008numerical,mitchell2014generalized}, and functional RG~\cite{Paris_2026}, establishing a unified description connecting these regimes and the universal crossovers between them remains a central challenge. This challenge has become particularly timely with recent experiments probing multichannel Kondo physics with unprecedented precision~\cite{iftikhar2015,iftikhar2018,piquard2023observing,piquard2026,pouse2023quantum,mitchell2016,karki2023z}.
	
	\begin{figure}
		\centering
		\includegraphics[width=\linewidth]{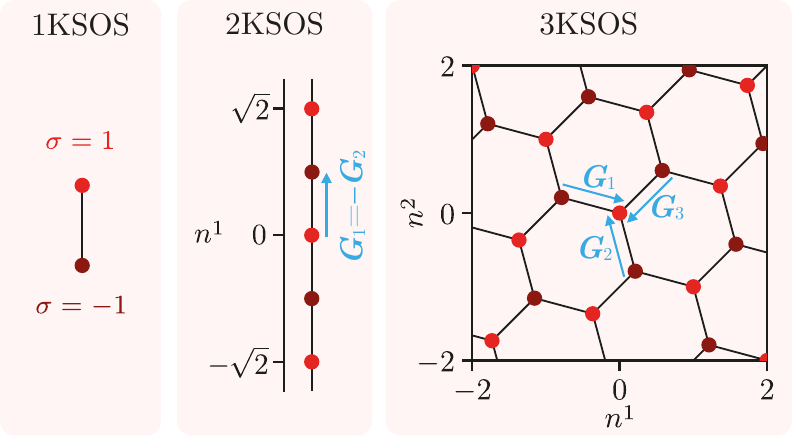}
		\caption{Configuration space of the fields $\sigma$ and $\bd n=(n^1,\dots,n^{N-1})$ appearing in the NKSOS model for $N=1,2,3$. The spin $\sigma$ distinguishes two Bravais lattices with $\sigma=1$ (red) and $\sigma=-1$ (brown). The reciprocal lattice vectors (in blue) are $\bd G_1,\dots,\bd G_N$.}
		\label{fig:3CK_potential}
	\end{figure}
	
	In this Letter, we introduce a discrete model that unifies the complementary descriptions of multichannel Kondo physics and provides direct access to the full crossovers between both the weak- and strong-tunneling regimes and the nontrivial multichannel Kondo fixed point. This model is inspired by the seminal Anderson-Yuval construction~\cite{anderson1969,Yuval_1970,anderson1970,Anderson_1971,benedetti2025} and can be simulated using highly efficient cluster Monte Carlo algorithms~\cite{Swendsen_1987,Wolff_cluster_algo,werner2005} adapted for long-range interactions \cite{Fukui_Todo_2009,ClockFMetManonMichel}. Remarkably, our algorithm does not suffer from any critical slowing down, making it several orders of magnitude faster than any standard Metropolis algorithm, and enabling simulations at very large system sizes and across a broad parameter range. Our results demonstrate that the seemingly distinct Kondo descriptions are manifestations of a single underlying universal structure. Finally, our approach naturally extends to large channel numbers, interacting one-dimensional leads, and asymmetric channel configurations. As a direct application, it yields predictions for transport in multichannel charge-Kondo mesoscopic devices~\cite{iftikhar2015,iftikhar2018}. Importantly, large channel numbers are experimentally accessible in these platforms, making the study of multichannel Kondo physics beyond the currently realized two- and three-channel cases an experimentally relevant direction.

	\paragraph{Kondo Solid-on-solid model ---}
	
	We introduce the $N$-channel Kondo solid-on-solid (NKSOS) model defined by the action $S=S_0+S_r$ with 
	\begin{align}\label{eq:action_S_0}
		S_0=&\frac{1}{2K}\sum_{\substack{i,j=1\\ (i\ne j)}}^\beta\frac{\left(\bd n_i-\bd n_j\right)^2 - J \sigma_i\sigma_j }{\left(\frac{\beta}{\pi}\right)^2 \sin^2\left(\frac{\pi(i-j)}{\beta}\right)} ,\\
		\label{eq:action_S_r}
		S_r=& r\sum_{i=1}^\beta \left[\left(\bd n_i-\bd n_{i+1}\right)^2 - J \sigma_i\sigma_{i+1}\right].
	\end{align}
	The lattice indexed by $i$ is a discretization of imaginary time where the lattice spacing (here set to 1) provides a UV cutoff similar to the bandwidth of an electron band. The inverse temperature $\beta$ is expressed in units of lattice spacing and is a dimensionless integer. At each site $i$, the field $\bd n_i = (n^1_i, \dots, n^{N-1}_i)$ takes its values among the coordinates of a $(N-1)$-dimensional hyperhoneycomb lattice with lattice vectors $\bd G_a$, $a\in \llbracket 1,N\rrbracket$, and the spin $\sigma_i=\sigma(\bd n_i)$ is $-1$ or $1$ depending on which Bravais sublattice $\bd n_i$ belongs to (see Fig.~\ref{fig:3CK_potential}). More formally, the allowed configurations $(\bd n_i,\sigma_i)$ are defined by a set of $N$ integers $M^a_i\in \mathbb{Z}$ such that $\bd n_i = \sum_a \bd G_a M^a_i$, $\sigma_i=2\sum_a M^a_i +1$ and with the constraint $\sum_a M^a_i=0$ or $-1$. The "Luttinger parameter" $K$ and isotropy parameter $J$ are linked to the Kondo physics (see next section), and $r$ controls the cost of each jump on the honeycomb lattice.
	
	The NKSOS model is an ideal setting to study the Kondo problem for two reasons. First, as we will demonstrate, it contains the exact same universal properties as the multichannel Kondo problem, and can thus be used to make quantitative predictions from high energies to low energies. Second, using cluster Monte Carlo algorithms adapted for long-range interactions~\cite{Swendsen_1987,Wolff_cluster_algo,Fukui_Todo_2009,ClockFMetManonMichel}, we are able to completely remove critical slowing down from our simulations and reach very large system sizes (see \cite{refSM} for details).
	
	For $N=1$, the field $\bd n$ disappears. Equations~\eqref{eq:action_S_0} and \eqref{eq:action_S_r} yield a long-range Ising model which is exactly the Anderson-Yuval approach to the $N=1$ Kondo model~\cite{anderson1969,Yuval_1970,anderson1970, Anderson_1971} and which has been studied extensively numerically in Refs.~\cite{Luijten_2001,Fukui_Todo_2009}.
	
	\emph{Link to the Kondo problem ---}
	The NKSOS model can be derived directly from the $N$-channel anisotropic Kondo Hamiltonian (NCK) $\hH=\hH_0+\hH_K$ with
	\begin{align}
		\hH_0&=i v_F\sum_{a=1}^N\sum_{\sigma=\uparrow,\downarrow} \int_{-\infty}^{+\infty}dx\,\hpsi_{a\sigma}^\dagger\partial_x\hpsi_{a\sigma},\\
		\hH_K&=v_F\sum_{a=1}^N \left[\lambda^z \hs_a^z(0) \hS_z+\frac{\lambda^\perp}{2}\left( \hs_a^{+}(0) \hS^- + \hs_a^{-}(0) \hS^+\right)\right],
	\end{align}
	where $\hpsi_{a\sigma}$ is a chiral fermion of spin $\sigma$ in channel $a$, $\hs_a^\mu (x) = \sum_{\sigma,\sigma'} \hpsi^\dagger_{a\sigma} (x) (\sigma^\mu_{\sigma,\sigma'}/2)\hpsi_{a\sigma'} (x)$ is the electronic spin density, and $\hS^\mu$ is the impurity spin. The NKSOS model is retrieved after bosonization and a Coulomb gas expansion (see~\cite{refSM} for more details). This procedure identifies $J=N/2[1/N - \lambda^z/(4\pi)]^2$, $r=-\log[\lambda^\perp/(4\pi)]/[1 -\lambda^z/(2\pi) + N{\lambda^z}^2/(16\pi^2)]$ and $K=1$. Varying $J$ between $0$ and $1/(2N)$ connects smoothly the Toulouse limit ($\lambda^z=4\pi/N$) to the planar Kondo model ($\lambda^z=0$). 
	
	The NKSOS model is also naturally connected to the highly transparent regime of charge-Kondo devices~\cite{Matveev_1995,furusaki_1995,iftikhar2018}. In that context, the spin of the Kondo model is replaced by a metallic island which contains $-M^a$ particles from channel $a$, and typically contains a total of $-\sum_a M^a=0$ or $1$ particle. In the NKSOS model, the variables $\bd n$ and $\sigma$ thus capture the distribution of the particles among the channels. The NKSOS model then describes the (imaginary-time) dynamics of the particles on the island (see~\cite{refSM} for more details). While the Kondo model strictly corresponds to the NKSOS model with $K=1$, adding interactions within the electronic leads of charge-Kondo devices can set $K\ne 1$ \cite{iftikhar2018,parafilo2024,ma2025}.
	
	The transport properties of the NKSOS model are directly related to the mobility~\cite{Yi_1998,Yi_2002} which quantifies the degree of localization of the fields $\bd n$ and $\sigma$.
	In charge Kondo experiments, the mobility maps directly onto the inter-channel linear conductance matrix $G_{ab}(i\omega,T=0)$. When all channels are identical, $G_{ab}(i\omega,T=0)=G(i\omega)\left(\delta_{ab}-\frac{1}{N}\right)$ with (see~\cite{Paris_2026,refSM})
	\begin{align}
		G(i\omega)=\frac{2\pi|\omega|}{(N-1) K}&\left[\langle|\bd n(i\omega)|^2\rangle + J/2\langle |\sigma(i\omega)|^2\rangle\right],
	\end{align}
	in units of $e^2K/h$. Its zero-frequency value is denoted by $G^*=G(0)$. Although our analysis hereafter focuses on the conductance, we emphasize that the NKSOS framework provides a general approach for computing equilibrium observables. As an additional example, we consider the susceptibility in Supplemental Material~\cite{refSM}.
	
	While the conductance depends on the microscopic parameter $r$, it exhibits a universal scaling behavior in the low-frequency limit $|\omega| \ll 1$ (recall that the UV cutoff is 1) as a function of $i\omega/T^*$,
	\begin{equation}
		G(i\omega)\underset{|\omega|\ll 1}{=}\mathcal{G}\left(\frac{i\omega}{T^*}\right),
	\end{equation}
	where $T^*$ is a $r$-dependent scale and $\mathcal{G}$ a universal function. In practice, we define $T^*$ such that $G(i\omega=T^*)$ lies halfway between its high- and low-frequency limits.

\begin{figure}[t]
		\centering
		\includegraphics[width=\linewidth]{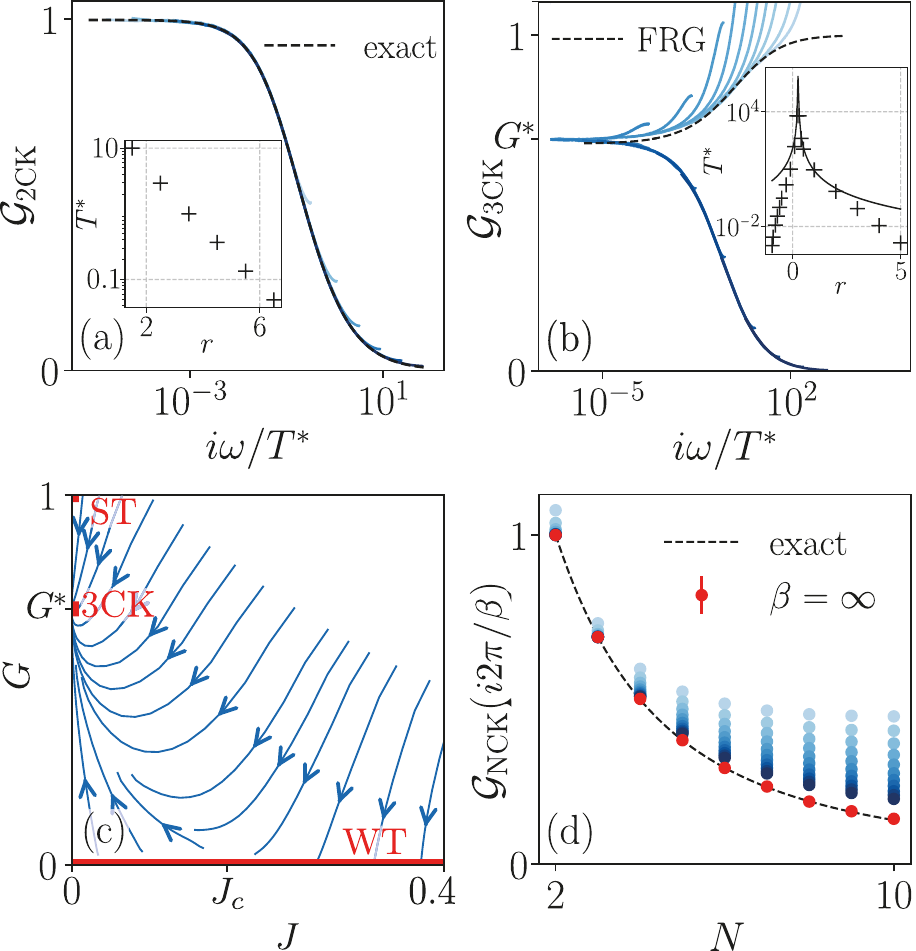}
		\caption{
			(a) In blue, conductance of the 2CK model for $J=0$ and for various values of $r$ in terms of the rescaled frequency. The dashed black line corresponds to the Emery-Kivelson solution~\eqref{eq:EK}. The crossover temperature is displayed in inset. 
			(b) Conductance of the 3CK model for $J=0$. The black dashed line is the FRG result obtained in Ref.~\cite{Paris_2026}. The $y$-tick $G^\ast$ indicates the exact value at zero frequency. For $r=4,5$, the values of the conductance at the two lowest frequencies are not plotted as they suffer from finite size effects. The crossover temperature $T^*$ is displayed in inset and diverges at a finite value $r_c\simeq 0.26$. Its behavior is compatible with the exact exponent given in Eq.~\eqref{eq:T_star_div} (solid black line).
			(c) RG flow in the $(J,G)$ plane for the 3KSOS model at $K=1$. The strong tunneling (ST) fixed point is unstable and flows to the 3CK fixed point. The weak tunneling (WT) line of fixed points is unstable for  $J \leq J_c=1/6$ and stable otherwise. The BKT transition is at $(G=0,J=J_c)$. For $N=2$, the NCK and ST fixed points would merge at $G=1$.
			(d) Conductance at frequency $2\pi/\beta$ for $\beta = 8,16,32,\dots,8192$ (blue) and extrapolated to $\beta=+\infty$ (red) for $K=1$ and $J=0$. The black dashed line represents the exact value $G^*_{N \rm CK}=2\sin^2(\pi/(N+2))$.
			Error bars are included in all the plots but are thinner than the lines and markers.}
		\label{fig:conductance_NCK}
	\end{figure}

	\paragraph{Anisotropic multi-channel Kondo model ---}
	In the two-channel Kondo model (2CK) at $J=0$, the conductance behaves as $G(0) = 1$ and $G(+\infty) = 0$, and the Emery-Kivelson solution~\cite{Emery_1992} provides an exact result for the universal crossover~\cite{refSM},
	\begin{equation}\label{eq:EK}
		\mathcal G\left(\frac{i\omega}{T^*}\right)=\frac{T^*\ln\left(1+| \omega|/T^*\right)}{| \omega|},
	\end{equation}
	with $T^*\propto e^{- r}$. 
	The universal scaling function obtained with our Monte-Carlo simulation of the KSOS model is in complete quantitative agreement with Eq.~\eqref{eq:EK}, as shown in Fig.~\ref{fig:conductance_NCK}(a). 
	
	The three-channel Kondo model (3CK) exhibits a more exotic behavior. The crossover temperature diverges at a finite value of the coupling $r=r_c$ as 
	\begin{equation}\label{eq:T_star_div}
		T^*\propto (r-r_c)^{-5/2},
	\end{equation}
	as shown in Ref.~\cite{iftikhar2018} and in agreement with experimental findings. This divergence separates the strong tunneling (ST) regime ($G(i\omega_n) \geq G^*_{\rm 3CK}$ in Fig.~\ref{fig:conductance_NCK}(b)) from the weak tunneling (WT) regime ($G(i\omega_n) \leq G^*_{\rm 3CK}$) that can both be explored experimentally~\cite{iftikhar2018}. Each regime has its own universal curve but both curve converge towards the same low-frequency value $G^*_{\rm 3CK}=2\sin^2(\pi/5)$. While the WT physics is well-captured by the numerical renormalization group (NRG) (see Refs.~\cite{iftikhar2018,mitchell2014}) and the ST one by the functional renormalization group (FRG)~\cite{Paris_2026}, so far no method has been able to describe both regimes or account for the divergence of $T^*$. Our method stands out as the first one able to do so. The zero-frequency value $G^*_{\rm 3CK}$ is recovered within $0.1\%$. The ST universal curve and the FRG predictions are consistent for $i\omega/T^*\gtrsim 1$ and slightly depart from each other at lower frequencies where the FRG is known to be less accurate.
	
	The phase diagram becomes considerably richer when $J$ is varied. The numerical results allow us to map out an effective RG flow, highlighting the behavior of the system across energy scales, as illustrated in Fig.~\ref{fig:conductance_NCK}(c) for the 3KSOS model. Inspired by Ref.~\cite{Luijten_2001}, we plot the conductance $G(i\omega)$ as a function of the scale-dependent renormalized coupling $J(j)=J\langle \sigma_i \sigma_{i+j}\rangle$~\cite{refSM}. Each line is a parametric curve $(G(i\omega_j),J(j))$ where we have paired the (lattice) times $j=1,2,4,8,\dots,\beta/2$ with the frequencies $\omega_j=\pi/j$ for $\beta=2^{17}=131072$. In this case, the WT phase is no longer represented by a single fixed point but by a continuous line of fixed points parameterized by $J$. A perturbative RG analysis shows that a Berezinskii--Kosterlitz--Thouless (BKT) transition separates stable WT fixed points with \mbox{$J \geq J_c=(K-1+1/N)/2$}, from unstable ones at $J\leq J_c$~\cite{Yi_2002}. Each unstable WT fixed point has its own RG trajectory connecting it to the 3CK fixed point. This implies that the WT to 3CK crossover is not described by a single, universal curve, but instead by a continuous family of crossover functions parametrized by the anisotropy. For the NKSOS model with $N\ne3$, a similar picture emerges with the WT to NCK crossover curve losing its uniqueness. For $N=2$ the 2CK and ST fixed points merge.
	
	As a result, collapsing Monte-Carlo or experimental data obtained at fixed $J$ and different values of $r$ does not reconstruct a single WT to NCK trajectory (except in the Toulouse limit $J=0$ which is stable under the RG) but rather combines bits of distinct flow lines.
	
	With most methods such as NRG or FRG calculations, considering more than three channels implies a huge numerical cost~\cite{iftikhar2018,Paris_2026}. Remarkably, our Monte Carlo simulations of the NKSOS model can accommodate more than $N=10$ channels, paving the way for experimentally relevant quantitative predictions of more complex Kondo systems. As a benchmark, the conductance at the NCK fixed point is computed up to $N=10$ and displayed in Fig.~\ref{fig:conductance_NCK}(d) (see~\cite{refSM} for details). The agreement with the exact value $G^*_{N \rm CK}=2\sin^2(\pi/(N+2))$ given in Refs.~\cite{Yi_1998,Yi_2002,bao_2017} is excellent~\footnote{The exact value is recovered within $1\%$ accuracy with moderate computational effort: a two-day simulation on a single computer for $N=10$.}.

\begin{figure}
		\centering
		\includegraphics[width=0.5\linewidth]{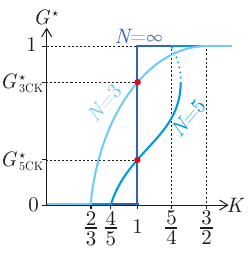}
		\includegraphics[width=0.48\linewidth]{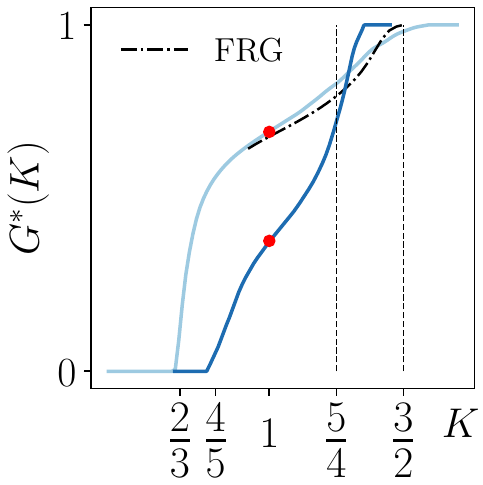}
		\caption{Left: Conductance as a function of $K$ as expected from perturbation theory and duality arguments~\cite{Yi_2002}. The first order transition appears for $N\ge 5$. Right: Conductance obtained from the NKSOS model with the multi-histogram method~\cite{ferrenberg1989} as a function of $K$ for the 3CK model (light blue) and the 5CK model (dark blue). The dashed and dotted black line indicates the FRG results from Ref.~\cite{Paris_2026} for the 3CK model.}
		\label{fig:cond_K}
	\end{figure}

	\paragraph{Kondo effect in interacting leads ---}
	Recently, experimental setups with interacting leads have triggered an increasing interest~\cite{anthore2018,parafilo2024,Paris_2026,ma2025,parafilo2022,parafilo2023,nguyen2020,Nguyen2020_2,karki_2024}. These correspond to the NKSOS model with a Luttinger parameter $K\ne 1$. In that case, there are no exact results from CFT or Bethe ansatz, but perturbation theory predicts a rich phase diagram, depending on the number of channels~\cite{Yi_1998,Yi_2002} (see Fig.~\ref{fig:cond_K} left). The ST fixed point ($G^*=1$) is stable for $K>N/(N-1)$ and the WT fixed point ($G^*=0$) for $K<(N-1)/N$. In the intermediate region, both are unstable, giving rise to a line of intermediate fixed points with $0<G^*<1$. 
	
	Our results confirm this scenario. The line of intermediate fixed points of the 3CK model is recovered from the extrapolation of finite temperature results, as displayed in Fig.~\ref{fig:cond_K}. Although extrapolating around $K=2/3$ and $K=3/2$ is made difficult by the presence of marginal operators, our results are consistent with perturbation theory in the vicinity of $K=2/3$, the exact result at $K=1$, and with recent FRG results~\cite{Paris_2026} for $K>1$. From a qualitative perspective, these intermediate fixed points exhibit Kondo-like physics: for any $K$, two universal conductance curves connecting $G=0$ and $G=1$ to an intermediate $K$-dependent $G^*$ can be found (similar to those shown in Fig.~\ref{fig:conductance_NCK}(b)).
	
	For $N\ge 5$, perturbation theory predicts that the line of intermediate fixed point continues beyond $K=N/(N-1)$ (see Fig.~\ref{fig:cond_K} left), together with the emergence of a reentrant line of unstable fixed points (dotted line in Fig.~\ref{fig:cond_K} left). This line signals a first-order transition associated with a discontinuous jump of the conductance. Numerically, approaching the line of fixed points from the weak-tunneling regime yields Fig.~\ref{fig:cond_K},  right. Although not quantitatively reliable in the region $K\simeq 5/4$ because of a marginal operator implying strong finite-size effects, our results fully endorse the existence of stable intermediate fixed points for $K>5/4$.
	
	In the limit where $N\to \infty$, the line of intermediate fixed points collapses onto a vertical line located at $K=1$, as depicted in Fig.~\ref{fig:cond_K} left. This can be understood intuitively by noting that, for finite $N$, the variable $\bd n = \sum_a M^a \bd G_a$ is constrained by $\sum_a M^a=0$ or $-1$. As $N\to \infty$, this constraint gets distributed over an infinite number of variables $M^a$ and becomes negligible for any individual degree of freedom. As shown formally in~\cite{refSM}, the resulting theory reduces to $N$ decoupled copies of the boundary sine-Gordon model. The $N=+\infty$ transition therefore belongs to the universality class of the Schmid transition \cite{Schmid_1983,Bulgadaev_1984}, which occurs at the single value $K=1$, as recently proven \cite{Paris_2025}.

	\paragraph{Nonequivalent channels ---} 
	The microscopic transparencies of the different channels are assumed identical in the action~(\ref{eq:action_S_r}). Although experimentally challenging, this assumption is crucial since any asymmetry is a relevant perturbation of the NCK fixed point~\cite{Yi_2002,iftikhar2018,bao_2017}. In the more general case where this symmetry is not satisfied, the NKSOS model becomes $S_0+S_{r_a}$ with
	\begin{align}\label{eq:action_SOS_asym}
		S_{r_a}=& \sum_{a=1}^N\sum_{i=1}^\beta  r_a\left[\left(\left(\bd n_i-\bd n_{i+1}\right)\cdot\bd G_a\right)^2 - \frac{J}{N} \sigma_i\sigma_{i+1}\right],
	\end{align}
	where $r_a$ controls the cost of a jump along the vector $\bd G_a$. For the sake of simplicity, we focus on the 3KSOS model at $J=0$ and introduce an asymmetry in the first channel, i.e. $r_2=r_3=r_{2,3}$ and $r_1\ne r_{2,3}$. This can produce a crossover from the 3CK fixed point to the 1CK or 2CK fixed points where, respectively, two and one channels totally decouple.
	
	\begin{figure}
		\centering
		\includegraphics[width=0.39\linewidth]{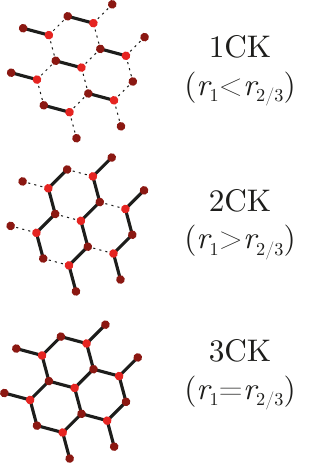}
		\includegraphics[width=0.59\linewidth]{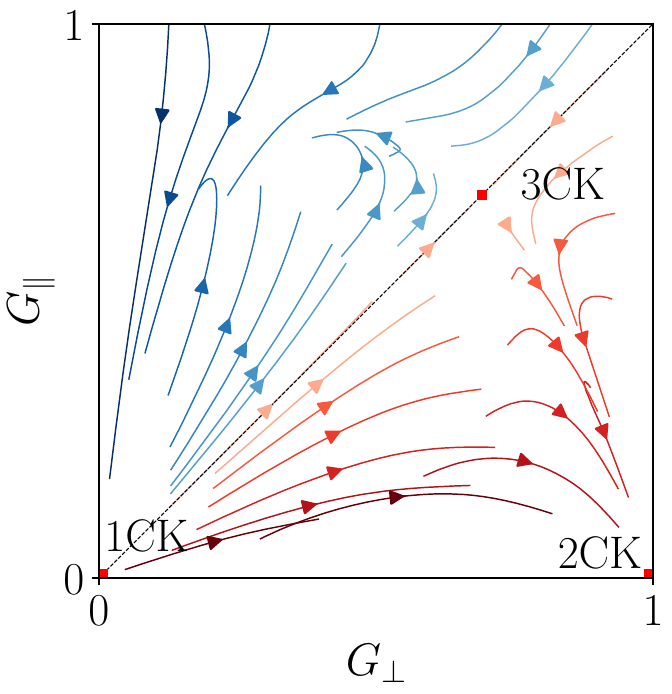}
		\caption{Left: schematic interpretation of the 1CK, 2CK and 3CK fixed points in the 3KSOS model. Right: 3KSOS RG flow towards the 1CK, 2CK and 3CK fixed points. The longitudinal conductance $G_\parallel(i\omega)$ is plotted as a function of the transverse conductance $G_\perp(i\omega)$.}
		\label{fig:asymmetric}
	\end{figure}
	
	The NKSOS model provides an intuitive explanation for these crossovers (see Fig.~\ref{fig:asymmetric}). When $r_1 < r_{2,3}$, jumps along the directions $\bd G_2$ and $\bd G_3$ become prohibitively costly at very low energies, so one expects the 3KSOS model to be described by an ensemble of isolated 1KSOS models (compare to Fig.~\ref{fig:3CK_potential}). This describes the 3CK to 1CK crossover. In the opposite case $r_1 > r_{2,3}$, the direction $\bd G_1$ is forbidden and we expect a crossover from the 3CK to the 2CK fixed point. On a more quantitative level, it is useful to introduce the longitudinal (i.e. along the first channel) conductance $G_\parallel(i\omega)= \frac{3 \pi|\omega|}{K}\langle |\bd G_1 \cdot \bd n(i\omega)|^2 \rangle$ and the transverse conductance $G_\perp(i\omega)=\frac{3 \pi|\omega|}{2K}\left[\langle |\bd G_2 \cdot \bd n(i\omega)|^2 \rangle+\langle |\bd G_3 \cdot \bd n(i\omega)|^2 \rangle\right]$, defined at $J=0$. In Fig.~\ref{fig:asymmetric}, following the procedure used in Fig.~\ref{fig:conductance_NCK}(c), we plot the parametric curves $(G_\perp(i\omega),G_\parallel(i\omega))$ for several microscopic conditions. Reading these curves from high to low frequencies reveals the RG flow connecting the 1CK, 2CK and 3CK fixed points similarly to the experimental findings of Ref.~\cite{iftikhar2015}.

	\paragraph{Conclusion ---}
	We have introduced a discrete model that extends the Anderson-Yuval construction to an arbitrary number of channels and can be connected to both the weak-tunneling Kondo model and the complementary strong tunneling of charge-Kondo devices. Using highly efficient cluster Monte Carlo algorithms, we showed that this single model captures the universal energy crossover from both weak and strong coupling to the nontrivial multichannel Kondo fixed point. Our results therefore unify descriptions previously obtained in complementary parameter regimes and from radically different approaches, providing evidence that these regimes belong to the same universality class. Owing to its flexibility with respect to channel number, channel asymmetry, and interactions in the leads, the present framework opens the door to quantitative studies of multichannel Kondo physics in regimes beyond the reach of existing approaches and provides a powerful tool for interpreting charge-Kondo experiments.

	\paragraph{Acknowledgements ---} We acknowledge fruitful discussions with N. Dupuis, A. Anthore, F. Pierre, C. Piquard, A. Mitchell and E. Sela. C.~M. acknowledges funding from the Agence Nationale de la Recherche under the France 2030 programme, reference ANR-22-PETQ-00122 (EQUBITFLY). O.B.-D. acknowledges the support of the French ANR under the grant ANR-22-CMAS-0001 (\emph{QuanTEdu-France} project).

\renewcommand{\addcontentsline}[3]{}

\let\addcontentsline\origaddcontentsline  

\clearpage
\newpage 

\onecolumngrid

\begingroup  

\setcounter{equation}{0}
\setcounter{figure}{0}
\setcounter{table}{0}
\setcounter{page}{1}
\renewcommand{\theequation}{S\arabic{equation}}
\renewcommand{\theHequation}{S\arabic{equation}}
\renewcommand{\thefigure}{S\arabic{figure}}	
\renewcommand{\theHfigure}{S\arabic{figure}}
\renewcommand{\thetable}{S\arabic{table}}	
\renewcommand{\theHsection}{S\arabic{section}}
\renewcommand{\theHsubsection}{S\arabic{section}.\Alph{subsection}}
\renewcommand{\theHsubsubsection}{S\arabic{section}.\Alph{subsection}.\arabic{subsubsection}}

\renewcommand{\bibnumfmt}[1]{[S#1]}
\renewcommand{\citenumfont}[1]{S#1}

\centerline{\bf Supplemental Material: Monte Carlo solution of the Kondo model}
\medskip
\centerline{Nicolas Paris, Oscar Bouverot-Dupuis, and Christophe Mora}

\bigskip 

\onecolumngrid

In this Supplemental Material, we explicitly derive the NKSOS model from the Kondo model and the charge Kondo model. We then detail the Monte Carlo algorithm used to simulate the NKSOS model and provide additional numerical results. Finally, we review the Emery--Kivelson solution for $N=2$ channels and analyze the large $N$ limit of the $N$-channel Kondo model.

\tableofcontents

\section{I. Conventions and notations}
The following list includes symbols and notations of frequent occurrence or special importance.\\\vspace{0.1cm}
\begin{tabular}{ l l }
	NKSOS & $N$-channel Kondo solid-on-solid model.\\ 
	NCK & $N$-channel Kondo model.\\ 
	QBM & quantum Brownian motion.\\
	$i,j,k$ & imaginary-time lattice site indices in $\llbracket 1, \beta \rrbracket$.\\
	$a,b,c$ & channel indices in $\llbracket 1, N \rrbracket$.\\  
	$\alpha,\beta$ & transverse channel indices in $\llbracket 1, N-1 \rrbracket$.\\
	$(\bd M, \bd \phi^\sigma, \bd \phi)=(M^a, \phi_\sigma^a, \phi^a)$ & $N$ channel variables in, respectively, the NKSOS, NCK and QBM models.\\
	$(\bd n, \bd \vartheta, \bd \varphi)=(n^\alpha, \vartheta^\alpha, \varphi^\alpha)$ & $N-1$ transverse channel variables in, respectively, the NKSOS, NCK and QBM models.\\
	$(\sigma,\vartheta^{\rm tot},\varphi^{\rm tot})$ & total channel variable in, respectively, the NKSOS, NCK and QBM models.\\
	$\bd G_a$ & vectors of $\mathbb{R}^{N-1}$ such that $\bd G_a \cdot \bd G_b=\delta_{ab}-1/N$.\\
	$\bd e_a$ & canonical basis of $\mathbb{R}^N$.\\
	$\phi^a(i\omega_n)=1/\sqrt{\beta} \int_0^\beta d \tau e^{-i\omega_n \tau}\phi^a(\tau)$ & continuous-time Fourier transform.\\
	$n^\alpha(i\omega_n)=1/\sqrt{\beta} \sum_{j=1}^\beta e^{-i\omega_n j}n^\alpha_j$ & discrete-time Fourier transform.
\end{tabular}

\section{II. Deriving the NKSOS model...}

A key finding of this work is that the NKSOS model is central to Kondo physics. In this Section, we establish mappings between the NKSOS and the various realizations of the Kondo model introduced in the literature, showing that they are all connected through the NKSOS as a common parent model. This unified framework naturally reveals the equivalence between different formulations of Kondo, which correspond to the weak- and strong-tunneling limits of the same underlying model. 
Figure~\ref{fig:Kondo_mappings} summarizes the relationships between the different Kondo models.

We begin in Section~\ref{sec:Nchannel_Kondo} by proving a rigorous mapping between the NKSOS and the original $N$-channel Kondo (NCK) model, given by
\begin{align}\label{eq:Kondo_Hamiltonian}
	\hH&=i v_F\sum_{a=1}^N\sum_{\sigma=\uparrow,\downarrow} \int_{-\infty}^{+\infty}dx\,\hpsi_{a\sigma}^\dagger\partial_x\hpsi_{a\sigma} + v_F\sum_{a=1}^N \left[\lambda^z \hs_a^z(0) \hS_z+\frac{\lambda^\perp}{2}\left( \hs_a^{+}(0)\hS^- + \hs_a^{-}(0)\hS^+\right)\right].
\end{align}
This equivalence holds independently of the anisotropy of the couplings. The relevant parameter range extends from the planar limit ($\lambda^z=0$) corresponding to purely in-plane antiferromagnetic coupling, through the isotropic Kondo point ($\lambda^z=\lambda^\perp$), up to the strongly anisotropic Toulouse point ($\lambda^z=4\pi/N$). Here, $\hpsi_{a\sigma}$ are $N$-chiral fermions with spin $\sigma$ and channel index $a$, and $\hs_a^\mu (x) = \sum_{\sigma,\sigma'} \hpsi^\dagger_{a\sigma} (x) (\sigma^\mu_{\sigma,\sigma'}/2)\hpsi_{a\sigma'} (x)$ is the local spin density with $\sigma^\mu$ Pauli matrices. After bosonizing the NCK model, we show that its Coulomb gas expansion exactly matches  the instanton expansion of the NKSOS.

In Section~\ref{sec:QBM}, we connect the NKSOS model to the $N$-channel charge Kondo model~\cite{Matveev_1995_sm,furusaki_1995_sm,Yi_1998_sm,Yi_2002_sm},
\begin{align}\label{eq:charge_Kondo}
	\hH = -iv_F \sum_{a=1}^N \int_{-\infty}^\infty dx \,(\hpsi_{a,R}^\dagger \partial_x\hpsi_{a,R} - \hpsi_{a,L}^\dagger \partial_x\hpsi_{a,L}) + E_C\left(\hat{N}-\frac{1}{2}\right)^2 + v_F \sum_{a=1}^N \tilde{r} [\hpsi_{a,R}^\dagger(0) \hpsi_{a,L}(0) + \hpsi_{a,L}^\dagger(0) \hpsi_{a,R}(0)],
\end{align}
which describes $N$ quantum Hall edge channels coupled to a central metallic island tuned to a charge degeneracy point (further details are provided in Sec.~\ref{sec:QBM}). This model is formulated using the ballistic limit ($\tilde{r}=0$) as a reference. By employing an instanton analysis in the limit of large $E_C$ and $\tilde{r}$, we rigorously demonstrate that the $N$-channel charge Kondo model maps onto the NKSOS model, and consequently, via the results of Sec.~\ref{sec:Nchannel_Kondo}, onto the standard Kondo model of Eq.~\eqref{eq:Kondo_Hamiltonian}. The ratio of these two large energy scales, $E_C/\tilde{r}$, dictates the degree of Kondo anisotropy $\lambda^z/\lambda^\perp$: the Toulouse limit is reached when $E_C \gg \tilde{r}$, whereas the planar Kondo limit ($\lambda^z=0$) corresponds to the opposite regime, $E_C \ll \tilde{r}$. Remarkably, even in the quasi-ballistic regime ($\tilde{r} \ll 1$) where the instanton approach is no longer formally justified, we find numerical evidence that the NKSOS model accurately reproduces the energy crossovers of the charge Kondo model~\eqref{eq:charge_Kondo} at large $E_C$. Finally, we note that for large $E_C$, Eq.~\eqref{eq:charge_Kondo} maps onto a quantum Brownian motion (QBM) model on a hyperhoneycomb lattice. Consequently, our framework also establishes a connection between the QBM and NKSOS models for arbitrary values of $\tilde{r}$.

The mappings we have established demonstrate that the full energy crossovers (from low to high energy) of the NKSOS model coincide with those of the two Kondo models, Eq.~\eqref{eq:Kondo_Hamiltonian} and Eq.~\eqref{eq:charge_Kondo}, provided that all energy scales are much smaller than the respective UV cutoffs. In other words, all three models belong to the same universality class. In the Kondo models, the conduction electron bandwidth plays the role of the UV cutoff, whereas for the NKSOS model, the UV cutoff is determined by the imaginary-time lattice spacing $\tau_c$ (taken equal to $1$ in our units). From a numerical perspective, the NKSOS model is the most convenient framework for exploring both the weak- and strong-tunneling regimes within a unified description.

\begin{figure}[h!]
	\centering
	\includegraphics[width=0.9\linewidth]{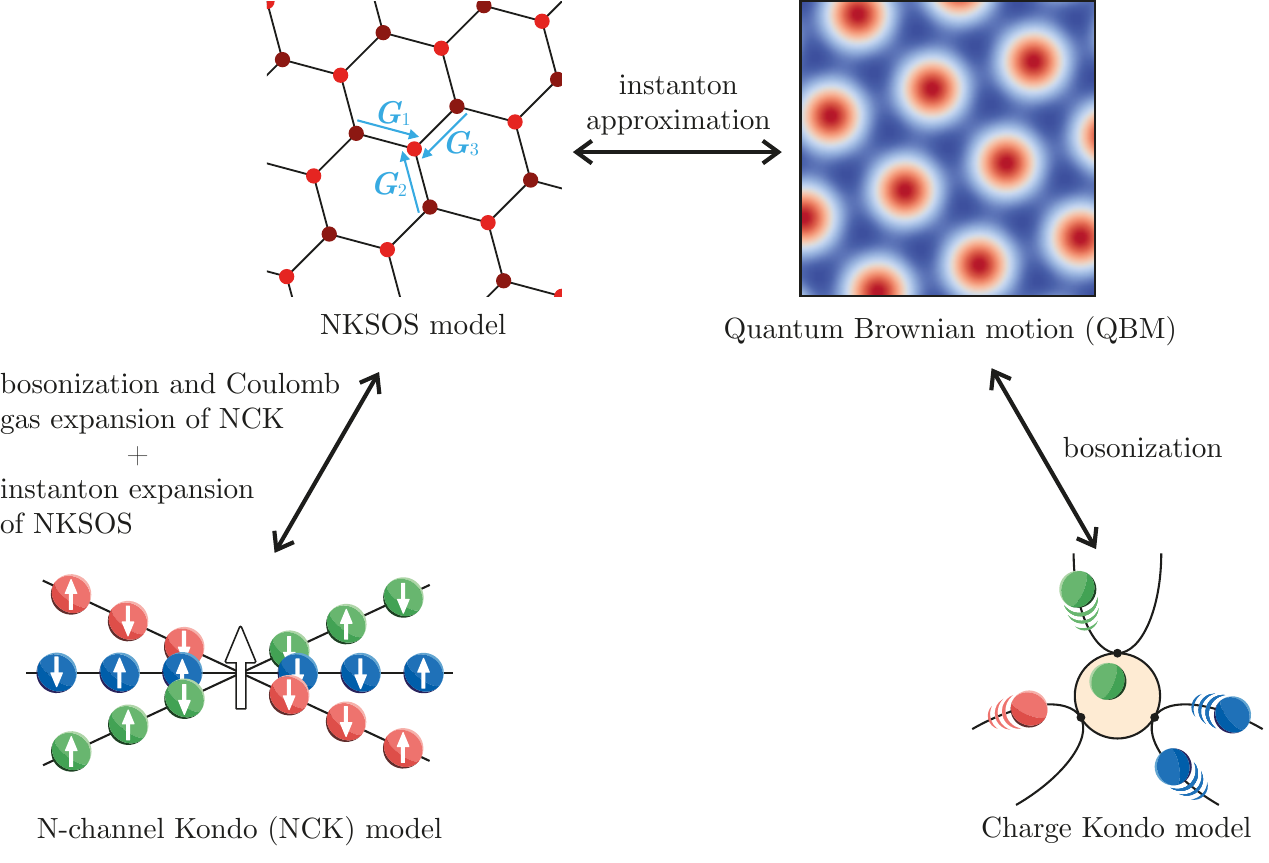}
	\caption{The NKSOS model (top left) can be derived from the $N$-channel Kondo model (bottom left) through bosonization and a Coulomb gas expansion for the NCK model and an instanton expansion for the NKSOS (Section~\ref{sec:Nchannel_Kondo}). It can also be derived from the charge Kondo model (bottom right) which, when bosonized, yields a quantum Brownian motion (QBM) model (top right) from which the NKSOS is recovered through an instanton analysis (Section~\ref{sec:QBM}).}
	\label{fig:Kondo_mappings}
\end{figure}

\subsection{A. ...from the $N$-channel Kondo model}
\label{sec:Nchannel_Kondo}
To map the NCK Hamiltonian onto the NKSOS model, we first bosonize the NCK model and then show that its Coulomb gas expansion exactly reproduces the instanton expansion of the NKSOS model.

\subsubsection{1. Bosonization of the Kondo model}
To make progress on the NCK model \eqref{eq:Kondo_Hamiltonian}, we first bosonize it using the identity~\cite{Emery_1992_sm}
\begin{equation}
	\hpsi_{a\sigma}(x)=\frac{\hU^a_\sigma}{\sqrt{2\pi \alpha}}e^{-i\hat{\Phi}^a_\sigma(x)},
\end{equation}
with $\hat{\Phi}^a_\sigma$ scalar fields such that $[\hat{\Phi}^a_\sigma(x),\hat{\Phi}^{a'}_{\sigma'}(x')]=-i\pi\delta_{aa'}\delta_{\sigma\sigma'}{\rm sign}(x-x')$, $\hU^a_\sigma$ Klein factors such that $\{\hU^a_\sigma,\hU^{a'}_{\sigma'}\}=2\delta_{aa'}\delta_{\sigma\sigma'}$ and $\alpha$ a short-distance cutoff coming from the electronic bandwidth. Further separating the charge modes $\hphi^a_\rho=\hat{\Phi}^a_\uparrow + \hat{\Phi}^a_\downarrow$ from the spin ones $\hphi^a_\sigma = \hat{\Phi}^a_\uparrow - \hat{\Phi}^a_\downarrow$, one arrives at the Hamiltonian $\hH=\hH_0+\hH_K$ with
\begin{align}
	\hH_0&=\frac{v_F}{8\pi}\sum_{a=1}^N\int dx\,(\partial_x\hphi^a_\sigma)^2 + \frac{v_F}{8\pi}\sum_{a=1}^N\int dx\,(\partial_x\hphi^a_\rho)^2,\\
	\hH_K&=\sum_{a=1}^{N}\left\{ \frac{v_F\lambda^z}{4\pi}\hS^z\partial_x\hphi^a_\sigma(0)+\frac{v_F\lambda^\perp}{4\pi \alpha}\left[\hS^- e^{i\hphi^a_\sigma(0)}+\hS^+ e^{-i\hphi^a_\sigma(0)}\right] \right\}.
\end{align}
The charge modes therefore decouple and are omitted in the following. Next, the $\hS_z\partial_x \hphi^a_\sigma(0)$ term is absorbed by applying the unitary transformation $\hU=e^{i\frac{\lambda^z}{4\pi} \hS^z\sum_a\hphi^a_\sigma(0)}$. This yields
\begin{align}
	\hH&=\sum_{a=1}^{N}\left\{\frac{v_F}{8\pi}\int_{-\infty}^{+\infty} dx\,(\partial_x\hphi^a_\sigma)^2+\frac{v_F\lambda^\perp}{4\pi \alpha}\left[\hS^- e^{i\hphi^a_\sigma(0)-i\frac{\lambda^z}{4\pi} \sum_b\hphi_\sigma^b(0)}+\hS^+ e^{-i\hphi^a_\sigma(0)+i\frac{\lambda^z}{4\pi} \sum_b\hphi^b_\sigma(0)}\right]\right\}.
\end{align}
It is useful to perform the orthonormal change of basis
\begin{equation}\label{eq:orthonormal}
	\begin{pmatrix}
		\hvartheta^{\rm tot}\\
		\hvartheta^1\\
		\vdots\\
		\hvartheta^{N-1}
	\end{pmatrix}=O
	\begin{pmatrix}
		\hphi^1_\sigma\\
		\vdots\\
		\hphi^N_\sigma
	\end{pmatrix}, \hspace{5mm} \text{with}\hspace{5mm} O=\begin{pmatrix}
		1/\sqrt{N}&\dots&1/\sqrt{N}\\
		\bd G_1&\dots&\bd G_N\\
	\end{pmatrix},
\end{equation}
to separate the total spin mode $\hvartheta^{\rm tot}$ from the channel-to-channel spin fluctuations $\bd \hvartheta=(\hvartheta^1,\dots,\hvartheta^{N-1})$ with commutation relations $[\hvartheta^\alpha(x),\hvartheta^\beta(x')]=-2i\pi \,{\rm sign}(x-x')\delta_{\alpha\beta}$. The vectors $\bd G_1,\dots,\bd G_N$ generate the $(N-1)$-dimensional regular simplex with normalization convention $\bd G_a\cdot \bd G_b=\delta_{ab} - 1/N$, and are defined up to a global rotation. A possible explicit construction is
\begin{equation}
	\begin{split}
		&\bd G_a=\bd e_a-\frac{1}{\sqrt{N}(\sqrt{N}+1)}(1,\dots,1), \hspace{5mm} {\rm for}\, a\in\{1,\dots,N-1\},\\
		&\bd G_{N}=-\frac{1}{\sqrt{N}}(1,\dots,1),
	\end{split}
\end{equation}
with $\{\bd e_a\}_{a\in\llbracket1,N-1\rrbracket}$ the canonical basis of $\mathbb R^{N-1}$. After this change of basis, the Hamiltonian comes down to
\begin{align}
	H=&\frac{v_F}{8\pi}\int_{-\infty}^{+\infty} dx\,\left[(\partial_x\hvartheta^{\rm tot})^2+(\partial_x\bd\hvartheta)^2\right]+t\sum_{a=1}^N\left[\hS^- \exp\left[i\bd G_a\cdot\bd\hvartheta(0)+i\gamma\hvartheta^{\rm tot}(0)\right]+{\rm H.c.}\right],
\end{align}
with
\begin{align}
	t=\frac{v_F\lambda^\perp}{4\pi \alpha}, \qquad \gamma=\sqrt{N}\left(\frac{1}{N}-\frac{\lambda^z}{4\pi}\right).
\end{align}
Switching to the action formalism for the phase variables $(\hvartheta^{\rm tot},\bd \hvartheta)$, the degrees of freedom at $x\ne 0$ can be integrated out \cite{Gogolin_2004_sm}. This leads to the Euclidean action
\begin{align}\label{eq:S_Kondo}
	&S_{0,\rm Kondo}=\frac{K}{4\pi}\sum_{\omega_n}|\omega_n|\left[|\vartheta^{\rm tot}(i\omega_n)|^2+|\bd\vartheta(i\omega_n)|^2\right],\\
	\label{eq:S_Kondo2}
	&S_{\rm int,\rm Kondo}=-t\int_{0}^{\beta} d\tau\sum_{a=1}^N\left[\sigma^- e^{i\bd G_a\cdot\bd\vartheta(\tau)+i\gamma\vartheta^{\rm tot}(\tau)}+{\rm H.c.}\right],
\end{align}
where $\sigma^\pm$ are Pauli matrices and $\vartheta^\alpha(\tau)$, $\vartheta^{\rm tot}(\tau)$ are scalar fields. We have also introduced a Luttinger parameter $K$ in Eq.~\eqref{eq:S_Kondo}. Although $K=1$ in the usual Kondo model, it can be varied in experimental realizations of the charge Kondo effect~\cite{parafilo2024_sm,ma2025_sm}. The sign of the coupling $t$ has no importance since it varies depending on the representation of the Pauli matrices.

A perturbative RG analysis in the weak-tunneling regime shows that the anisotropic Kondo model exhibits a Berezinskii--Kosterlitz--Thouless (BKT) transition, characterized by the standard separatrix structure in the $(\gamma,t)$ plane~\cite{Yi_1998_sm,Yi_2002_sm}:
\begin{align}
	\frac{dt}{dl}&=t\left[1-\frac{1}{K}\left(1-\frac{1}{N}+\gamma^2\right)\right]\\
	\frac{d\gamma}{dl}&= -\frac{t^2\gamma}{2}
\end{align}
where the factor $1/2$ in the second line is non-universal. The same BKT transition is visible in the RG flow of the KSOS model shown in Fig.~2(c) of the main text~\footnote{The two models belong to the same universality class. In the weak-tunneling regime, where the transition occurs, the conductance is related to the tunneling amplitude as $G\propto t^2$ via a perturbative expansion, while the KSOS coupling $J$ is related to the anisotropy $\gamma$ through Eq.~\eqref{eq:Delta_Kondo}.}.

\subsubsection{2. Coulomb gas expansion of the Kondo model\label{sec:Coulomb_gas}}

The partition function associated with the action~\eqref{eq:S_Kondo} is
\begin{align}
	Z&={\rm Tr_{\sigma}}\int \mathcal{D}\bd \vartheta\mathcal{D}\vartheta^{\rm tot} \, e^{-S_{0,\rm Kondo}-S_{\rm int,\rm Kondo}},
\end{align}
where we trace over the 2-state Hilbert space of the Pauli matrices since we have not used the action formalism for the impurity degree of freedom. Performing an expression in powers of $t$ yields
\begin{align}
	Z&={\rm Tr_{\sigma}}\sum_m \frac{t^m}{m!}\left\langle\left(\sum_{a=1}^N\int_{0}^{\beta} d\tau[\sigma^+e^{-i\bd G_a\cdot \bd\vartheta(\tau)-i\gamma\vartheta^{\rm tot}(\tau)}+{\rm H.c.}]\right)^m\right\rangle_0\\
	&=2\sum_{m \, \rm even}t^m\int_{\tau_1<\dots<\tau_m}\sum_{\{a_p\}}\left\langle  e^{-i\sum_{p=1}^m\left[(-1)^p\bd G_{a_p}\cdot \bd \vartheta(\tau_p)+(-1)^p\gamma\vartheta^{\rm tot}(\tau_p)\right]}\right\rangle_0\\
	&=2\sum_{m \, \rm even}t^m\int_{\tau_1<\dots<\tau_m}\sum_{\{ a_p\}}\delta_{\sum_{p=1}^m (-1)^p\bd G_{a_p},0}\exp\left(\frac{2}{K }\sum_{p<q}(-1)^{p+q}(\bd G_{a_p} \cdot \bd G_{a_q}+\gamma^2)\log\left(\frac{\beta}{\pi \tau_c}\sin\left(\pi(\tau_p-\tau_q)/\beta\right)\right)\right),\label{eq:Coulomb_gas}
\end{align}
where the average $\langle\cdot\rangle_0$ is taken with respect to $S_{0,\rm Kondo}$ and $\sum_{\{a_p\}}=\prod_{p=1}^m \sum_{a_p=1}^N$. From the first line to the second we have used the fact that only the alternating matrix products $\sigma^+\sigma^-\sigma^+\sigma^-\dots$ and $\sigma^-\sigma^+\sigma^-\sigma^+\dots$ are non-zero. Going to the last line, we split the fields as $\vartheta^a(\tau)=\vartheta_0^a + \tilde \vartheta^a(\tau)$, for $a={\rm tot},1,\dots, N-1$, where $\vartheta_0^a$ is the zero-frequency mode and $\tilde \vartheta^a(\tau)$ contains only non-zero modes. The Kronecker delta $\delta_{\sum_p (-1)^p\bd G_{a_p},0}$ comes from integrating out $\vartheta_0^a$, while the long range interactions come from $\langle [\tilde \vartheta^a(\tau) - \tilde \vartheta^a(\tau')]^2\rangle_0 = 4/K\log\left(\frac{\beta}{\pi \tau_c}\sin\left(\pi(\tau-\tau')/\beta\right)\right)$ with $\tau_c = \alpha/v_F$ the short-time cutoff.

\subsubsection{3. Instanton expansion of the NKSOS model}
We now perform an instanton expansion of the NKSOS model and show that it reproduces the previous Coulomb gas expansion of the NCK model. The partition function of the NKSOS model is given by 
\begin{equation}
	Z_{\rm KSOS}=\int \mathcal D\bd n\mathcal D\sigma \, e^{-S_0-S_r},
\end{equation}
with the action 
\begin{align}
	&S_0=\frac{1}{2K}\sum_{\substack{i,j=1\\(i\ne j)}}^{\tilde\beta} \frac{(\bd n_i-\bd n_j)^2 - J \sigma_i\sigma_j}{\left(\frac{\tilde\beta}{\pi}\right)^2\sin^2\left(\frac{\pi(i-j)}{\tilde\beta}\right)},\\
	&S_r= r\sum_{i=1}^{\tilde\beta} \left[\left(\bd n_i-\bd n_{i+1}\right)^2 - J \sigma_i\sigma_{i+1}\right].
\end{align}
In this section, we denote the inverse temperature in units of the lattice spacing by $\tilde \beta$ to avoid confusion with the dimensionful $\beta$ of the previous paragraph.  We emphasize that we do not make this distinction explicit and denote it by $\beta$ in the other sections.
The configurations $\bd n$, $\sigma$ that contribute to the partition function are periodic, i.e. $\bd n_{\tilde\beta}=\bd n_0$ and $\sigma_{\tilde\beta}=\sigma_0$, and can be written explicitly as the sum of consecutive jumps, in the spirit of an instanton expansion~\cite{NDbook1_sm}. Due to the properties of the $(N-1)$-dimensional honeycomb lattice, these jumps alternate between jumps along the vectors $+G_a$ and the vectors $-G_a$ (see Fig.~\ref{fig:3CK_Toulouse_QBM} right). If the first jump is along $-G_a$, one can write
\begin{align}\label{eq:instanton1}
	\partial_{\tilde \tau} \bd n(\tilde \tau)&=\sum_{p=1}^m (-1)^p \bd G_{a_p}\delta(\tilde \tau-\tilde \tau_p),\\
	\label{eq;instanton2}
	\partial_{\tilde \tau} \sigma(\tilde \tau)&=2\sum_{p=1}^m (-1)^p \delta(\tilde \tau - \tilde \tau_p),
\end{align}
or equivalently in Fourier space
\begin{align}\label{eq:instanton_Fourier1}
	\bd n(i\tilde \omega_n)&=\frac{1}{i\tilde \omega_n\sqrt{\tilde\beta}}\sum_{p=1}^m (-1)^p \bd G_{a_p} e^{-i\tilde \omega_n\tilde \tau_p},\\
	\label{eq:instanton_Fourier2}
	\sigma(i\tilde \omega_n)&=\frac{2}{i\tilde \omega_n\sqrt{\tilde\beta}}\sum_{p=1}^m (-1)^p e^{-i\tilde \omega_n\tilde \tau_p}.
\end{align}
We have switched from the lattice representation to a continuous  $\tilde \tau$ with an implicit UV cutoff set by the lattice spacing. Since this spacing is unity, the  $\tilde \tau$ and Matsubara frequencies $\tilde \omega_n$ are dimensionless, hence their names. $\delta$ is the Dirac delta, $\tilde \tau_1<\dots<\tilde \tau_m$ are the times of the $m$ jumps forming the instanton trajectory, and $\{a_p\}_{p\in\llbracket 1,n\rrbracket}$ lists the jump directions. The discrete or continuous nature of the imaginary time does not affect the universal properties which appear at frequencies and temperatures far below the (energy) UV cut-off.

One further notices that the periodicity in imaginary time imposes 
\begin{align}\label{eq:period_cond}
	\sum_{p=1}^m (-1)^p=0 \Leftrightarrow m \text{ even}, \qquad \sum_{p=1}^m (-1)^p \bd G_{a_p} = 0.
\end{align}
Evaluating the action on the previous continuous-time ansatz is easily done once one realizes the action $S_r$ weights uniformly the $m$ instantons. This yields
\begin{align}
	S_0 + S_r =&  \frac{1}{2K}\int_0^{\tilde\beta} d \tilde \tau d \tilde \tau'  \frac{[\bd n(\tilde \tau)-\bd n(\tilde \tau')]^2 + J/2 [\sigma(\tilde \tau)-\sigma(\tilde \tau')]^2}{\left(\frac{\tilde\beta}{\pi}\right)^2\sin^2\left(\frac{\pi(\tilde \tau - \tilde\tau')}{\tilde\beta}\right)}  + \sum_{p=1}^m r \left( \bd G_{a_p}^2 + 2J\right)\\
	=& \frac{\pi}{K} \sum_{p,q=1}^m (-1)^{p+q}(\bd G_{a_p}\cdot \bd G_{a_q}  + 2J ) \frac{1}{\tilde\beta} \sum_{\tilde \omega_n} \frac{e^{i\tilde\omega_n(\tilde\tau_p - \tilde\tau_q)} - 1}{|\tilde\omega_n|}  + m r \left( 1 - \frac{1}{N} + 2J\right)\\
	=& - \frac{2}{K}\sum_{p<q}(-1)^{p+q} (\bd G_{a_p} \cdot \bd G_{a_q} + 2J)\log\left(\tilde \beta/\pi\sin\left(\pi(\tilde\tau_p-\tilde\tau_q)/\tilde\beta\right)\right) + m r \left( 1- \frac{1}{N} + 2J\right).
\end{align}
From the first to the second line we have used the Fourier transformed ansatz (\ref{eq:instanton_Fourier1},\ref{eq:instanton_Fourier2}), the fact that the Fourier transform of $1/[(\tilde\beta/\pi)^2\sin^2(\pi \tilde\tau/\tilde\beta)]$ is $-\pi |\tilde\omega_n|$, and the conditions~\eqref{eq:period_cond} to add $-\pi/(K\beta) \sum_{\tilde\omega_n}1/|\tilde\omega_n| \sum_{p,q}(-1)^{p+q} (\bd G_{a_p} \cdot \bd G_{a_q} + 2J)$. The third line is then obtained by performing the Matsubara sum. The partition function is finally found by summing over all valid configurations,
\begin{align}
	Z_{\rm KSOS}=&2 \sum_{m\, even} e^{-mr\left(1-\frac{1}{N}+2J\right)}\int_{\tilde\tau_1<\dots<\tilde\tau_m} \sum_{\{a_p\}} \delta_{\sum_{p=1}^m  (-1)^p \bd G_{a_p},0}\nonumber\\
	&\times\exp\left(\frac{2}{K}\sum_{p<q}(-1)^{p+q} (\bd G_{a_p} \cdot \bd G_{a_q} + 2J)\log\left(\tilde \beta\sin\left(\pi(\tilde\tau_p-\tilde\tau_q)/\tilde\beta\right)\right) \right),
\end{align}
and the factor of $2$ accounts for the configurations starting with a jump along $+\bd G_a$ instead of $-\bd G_a$. This is precisely the Coulomb gas expansion of the NCK model~\eqref{eq:Coulomb_gas} provided one identifies
\begin{align}
	&\tilde\beta= \frac{\beta}{\tau_c},\\
	&\tilde \tau= \frac{\tau}{\tau_c},\\
	\label{eq:r_Kondo}
	&r\left(1-\frac{1}{N} + 2J\right)=-\log\left(t\tau_c\right)=-\log\left(\frac{\lambda^\perp}{4\pi}\right),\\
	\label{eq:Delta_Kondo}
	&J=\frac{\gamma^2}{2}=\frac{N}{2}\left(\frac{1}{N}-\frac{\lambda^z}{4\pi}\right)^2.
\end{align}
The arguments shown above straightforwardly generalize to the case of inequivalent channels. The NCK model with a channel-dependent coupling $\lambda_a^\perp$ gets mapped onto the modified NKSOS model $S=S_0 + S_{r_a}$ with
\begin{align}
	S_{r_a}= \sum_{a=1}^N r_a \sum_{i=1}^{\tilde\beta} \left[\left(\bd G_a \cdot (\bd n_i-\bd n_{i+1})\right)^2 - \frac{J}{N} \sigma_i \sigma_{i+1}\right].
\end{align}
provided Eq.~\eqref{eq:r_Kondo} is replaced by
\begin{align}
	&r_a\left(1-\frac2N\right) + \sum_{b=1}^N r_b \left(\frac{1}{N^2} + \frac{2J}{N}\right)=-\log\left(\frac{\lambda_a^\perp}{4\pi}\right).
\end{align}

\subsection{B. ...from the charge Kondo model}
\label{sec:QBM}
In this section, we argue that the NKSOS model is also related to the charge Kondo model~\eqref{eq:charge_Kondo}. We rewrite the Hamiltonian for clarity:
\begin{align}\label{eq:charge_Kondo2}
	\hH = -iv_F \sum_{a=1}^N \int_{-\infty}^\infty dx\, (\hpsi_{a,R}^\dagger \partial_x\hpsi_{a,R} - \hpsi_{a,L}^\dagger \partial_x\hpsi_{a,L}) + E_C\left(\hat{N}-\frac{1}{2}\right)^2 + v_F \sum_{a=1}^N \tilde{r} [\hpsi_{a,R}^\dagger(0) \hpsi_{a,L}(0) + \hpsi_{a,L}^\dagger(0) \hpsi_{a,R}(0)],
\end{align}
which describes $N$ quantum Hall edge channels coupled through quantum point contacts (QPCs) to a metallic island with charging energy $E_C$, as illustrated in Fig.~\ref{fig:Kondo_mappings} and realized experimentally in Ref.~\cite{iftikhar2018_sm}. Here, $\hpsi_{a,R/L}$ denote right- and left-moving fermionic fields in channel $a$, with $x>0$ corresponding to the region outside the island and $x<0$ to its interior. The QPCs, located at $x=0$, backscatter electrons with amplitude $\tilde r$. A gate voltage is tuned to the charge-degeneracy point, such that the electrostatically preferred island charge is $1/2$.

Following Ref.~\cite{Paris_2026_sm}, bosonization and integration over the lead degrees of freedom outside $x=0$ map Eq.~\eqref{eq:charge_Kondo2} onto the quantum Brownian motion of a particle moving in a $(N-1)$-dimensional potential,
\begin{align}\label{eq:action_charge}
	S_{\rm QBM}=\sum_{a=1}^N \sum_{\omega_n}\frac{|\omega_n|}{\pi K}|\phi^a(i\omega_n)|^2 - \sum_{a=1}^N \tilde r \int _0^{\beta}d\tau \cos(2\phi^a(\tau))+ E_C\int _0^\beta d\tau\biggl(\frac{1}{\pi}\sum_{a=1}^N\phi^a(\tau)+\frac{1}{2}\biggr)^2,
\end{align}
where unimportant numerical prefactors have been absorbed into $\tilde r$. The parameter $K$ denotes the Luttinger parameter of the one-dimensional electronic channels. While $K=1$ in the charge Kondo model~\eqref{eq:charge_Kondo}, it can be tuned experimentally, for instance by introducing additional islands in the setup~\cite{parafilo2024_sm,anthore2018_sm}. We first explicitly show that the limit $E_C\gg \tilde r$ corresponds to the NKSOS model at $J=0$, which is the Toulouse limit of the Kondo model ($\lambda^z=4\pi/N$), and that $E_C \ll \tilde r$ corresponds to the NKSOS model at $J=1/(2N)$, which is the planar Kondo model ($\lambda^z=0$). We then argue that the NKSOS model captures the full crossover between $E_C$ and $\tilde r$. 

The derivation of these mappings relies on an instanton expansion of the QBM model and is therefore formally controlled in the regime of large $\tilde r$ and $E_C$, which corresponds to the weak-tunneling regime. Nevertheless, the numerical results presented in the main text demonstrate that the NKSOS model correctly reproduces the Kondo universality for all regimes of parameters. In particular, this includes the quasi-ballistic (strong-tunneling) regime $\tilde r\ll 1$.

\subsubsection{1. Toulouse limit}
When the charging energy $E_C$ is very large, the total charge $-\frac{1}{\pi}\sum_a \phi^a$ in Eq.~\eqref{eq:action_charge} gets stuck around $1/2$. To isolate this charge mode $\varphi^{\rm tot}=\frac{1}{\sqrt{N}}\sum_a\phi^a$ from the channel-to-channel fluctuations $\bd \varphi=(\varphi^1,\dots,\varphi^{N-1})$ we perform the orthonormal transformation~\eqref{eq:orthonormal}
\begin{equation}
	\begin{pmatrix}
		\varphi^{\rm tot}\\
		\varphi^1\\
		\vdots\\
		\varphi^{N-1}
	\end{pmatrix}=O
	\begin{pmatrix}
		\phi^1\\
		\vdots\\
		\phi^{N}
	\end{pmatrix},
\end{equation}
which leads to the action
\begin{align}
	S = \sum_{\omega_n} \frac{|\omega_n|}{\pi K}\left[|\bd \varphi(i\omega_n)|^2+|\varphi^{\rm tot}(i\omega_n)|^2\right]-r\int_0^\beta d\tau\sum_{a=1}^{N}\cos \left(2\bd G_a\cdot \bd \varphi(\tau)+\frac{2}{\sqrt{N}}\varphi^{\rm tot}\right)+E_C\int d\tau\left(\frac{\sqrt{N}}{\pi}\varphi^{\rm tot}+\frac{1}{2}\right)^2.
\end{align}
Taking $E_C\to \infty$ then amounts to replacing $\varphi^{\rm tot} \to -\pi/(2\sqrt{N})$. This yields an effective action for $\bd \varphi$ corresponding to a quantum Brownian motion in a potential periodic over a hypertriangular lattice with $\pi/N$ flux per plaquette~\cite{Yi_2002_sm}. 
In real space it reads
\begin{align}\label{eq:QBM_toulouse}
	S=\frac{1}{2\pi^2 K} \int_0^\beta d \tau d \tau' \frac{[\bd \varphi(\tau)-\bd \varphi(\tau')]^2}{\left(\frac{\beta}{\pi}\right)^2\sin^2\left(\frac{\pi(\tau - \tau')}{\beta}\right)} - \tilde r\sum_{a=1}^{N}\int_0^\beta d\tau\cos \left(2\bd G_a\cdot \bd \varphi(\tau)-\frac{\pi}{N}\right).
\end{align}
In the limit where $\tilde r$ is large (but $\tilde r \ll E_C$), the field $\bd \varphi$ gets localized in the wells of the periodic potential $V(\bd \varphi)=-\sum_a \cos(2\bd G_a\cdot \bd \varphi - \frac{\pi}{N})$ and moves between neighbouring minima via instantons. These minima are located at $\bd\varphi=\pi \bd n$ with $\bd n=\sum_a M_a\bd G_a$, with $M_a\in\mathbb Z$ satisfying the constraint $\sum_a M_a\in\{-1,0\}$. They form a hyperhoneycomb lattice generated by the vectors $\bd G_a$ (see Fig.~\ref{fig:3CK_Toulouse_QBM}). Intuitively, one can thus replace $\bd\varphi$ by the discrete field $\pi\bd n$ living on the minima of the potential and introduce a parameter $r$ to account for the cost of instantons. After discretizing the imaginary time at the cutoff scale $\tau_c$ and going to units where $\tau_c=1$, this leads to 
\begin{align}
	S=&\frac{1}{2K} \sum_{\substack{i,j=1\\(i\ne j)}}^\beta \frac{(\bd n_i-\bd n_j)^2}{\left(\frac{\beta}{\pi}\right)^2\sin^2\left(\frac{\pi(i-j)}{\beta}\right)} + r\sum_{i=1}^\beta \left(\bd n_i-\bd n_{i+1}\right)^2,
\end{align}
which is nothing but the NKSOS model at $J=0$. From Eq.~\eqref{eq:Delta_Kondo}, $J=0$ corresponds to the NCK model with $\lambda^z=4\pi/N$, which is the Toulouse limit of the Kondo model \cite{Yi_2002_sm}.
\begin{figure}[h]
	\centering
	\includegraphics[width=0.7\linewidth]{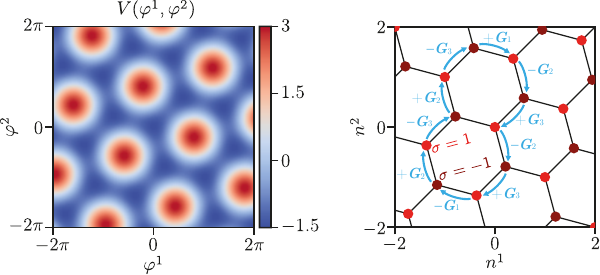}
	\caption{Left: potential $V(\bd \varphi)=-\sum_a \cos(2\bd G_a\cdot \bd \varphi - \frac{\pi}{N})$ for $N=3$. Right: the lattice of minima of $V(\bd \varphi)$ forms the configuration space of the NKSOS model. Jumps between minima correspond to instantons. We have represented in blue a cycle of 8 instantons.}
	\label{fig:3CK_Toulouse_QBM}
\end{figure}

\subsubsection{2. Planar Kondo limit}
In the opposite regime where $\tilde r$ is much larger than $E_C$, the action~\eqref{eq:action_charge} can be directly expanded in instantons of the potential $V(\bd \phi)=- \sum_a \cos(2\phi^a)$ and one replaces $\phi^a\to \pi M^a$ with $M^a \in \mathbb{Z}$ (see Fig.~\ref{fig:2CK_Planar_QBM}). We then consider $E_C$ large (but $E_C\ll \tilde r$) which restricts $\sum_a M_a$ to be either $0$ or $-1$. From a geometrical point of view, $\bd M=(M^1,\dots,M^N)$ lives on a hypercubic lattice (see Fig.~\ref{fig:2CK_Planar_QBM}). Going through the---now familiar---steps of discretizing imaginary time, making it dimensionless and adding a cost to instantons gives
\begin{align}\label{eq:KSOS_N}
	S=&\frac{1}{2K} \sum_{\substack{i,j=1\\(i\ne j)}}^\beta \frac{(\bd M_i-\bd M_j)^2}{\left(\frac{\beta}{\pi}\right)^2\sin^2\left(\frac{\pi(i-j)}{\beta}\right)} + r\sum_{i=1}^\beta \left(\bd M_i-\bd M_{i+1}\right)^2.
\end{align}
Introducing the variables $\sigma$, $\bd n$ through the orthogonal transformation~\eqref{eq:orthonormal}
\begin{equation}
	\begin{pmatrix}
		(\sigma-1)/(2\sqrt{N})\\
		n^1\\
		\vdots\\
		n^{N-1}
	\end{pmatrix}=O\begin{pmatrix}
		M^1\\
		\vdots\\
		M^N
	\end{pmatrix},
\end{equation}
the action \eqref{eq:KSOS_N} becomes,
\begin{align}\label{eq:KSOS_planar}
	S=&\frac{1}{2K} \sum_{\substack{i,j=1\\(i\ne j)}}^\beta \frac{(\bd n_i-\bd n_j)^2 - \frac{1}{2N}\sigma_i\sigma_j}{\left(\frac{\beta}{\pi}\right)^2\sin^2\left(\frac{\pi(i-j)}{\beta}\right)} + r\sum_{i=1}^\beta \left[(\bd n_i-\bd n_{i+1})^2 - \frac{1}{2N}\sigma_i\sigma_{i+1}\right],
\end{align}
The vector $\bd n$ takes its values on a hyperhoneycomb lattice split into two sublattices with $\sigma=-1$ and $\sigma=1$ depicted in red and brown in Figure~\ref{fig:2CK_Planar_QBM}. Therefore, Eq.~\eqref{eq:KSOS_planar} is exactly the NKSOS model for $J=1/(2N)$. Looking back at Eq.~\eqref{eq:Delta_Kondo}, $J=1/(2N)$ corresponds to the Kondo model with $\lambda^z=0$, also known as the planar Kondo model. 
\begin{figure}[h]
	\centering
	\includegraphics[width=0.7\linewidth]{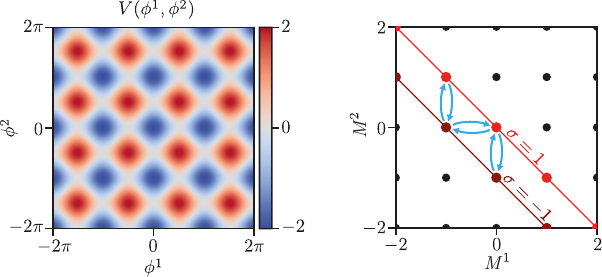}
	\caption{Left: potential $V(\bd \phi)=-\sum_a \cos(2 \phi^a)$ for $N=2$. Right: the lattice of minima of $V(\bd \phi)$ forms the configuration space of the NKSOS model. The jumps between minima going downwards and right correspond to $+\bd G_1$ instantons, while the upwards and left ones correspond to $\bd G_2=-\bd G_1$ instantons. In blue is a cycle of 6 instantons.}
	\label{fig:2CK_Planar_QBM}
\end{figure}

\subsubsection{3. General case}
For a finite ratio $E_C/\tilde r$, the possible instantons are found by analyzing the minima of the full QBM potential
\begin{equation}
	V(\bd \varphi,\varphi^{\rm tot})=-\tilde r \sum_a \cos\left(2\bd G_a\cdot \bd\varphi + \frac{2}{\sqrt{N}} \varphi^{\rm tot}\right) + E_C\left(\frac{\sqrt{N}}{\pi} \varphi^{\rm tot} +1/2\right)^2.
\end{equation}
These lie at
\begin{align}
	\label{eq:varphi_Ec_r}
	\bd \varphi=&\pi \bd n=\pi \sum_a \bd G_a M^a \text{ with } M^a\in \mathbb{Z} \text{ and } \sigma=2\sum_a M_a+1 \in \{-1, 1\},\\
	\label{eq:varphi_c_Ec_r}
	\varphi^{\rm tot}=&\varphi^{\rm tot}(\sigma)=\frac{\pi}{2}\left[\sqrt{2J}\sigma - 1/\sqrt{N}\right],
\end{align}
with $J$ such that
\begin{equation}
	\frac{\partial V}{\partial \varphi^{\rm tot}}(\bd \varphi=\pi \bd n,\varphi^{\rm tot}=\varphi^{\rm tot}(\sigma))=0 \Rightarrow \tilde r \sin\left(\frac{\pi}{N}(\sqrt{2J N}-1)\right) + \frac{E_C}{2 } \sqrt{2J N}=0.
\end{equation}
This equation cannot be analytically solved but it interpolates between $J=0$ for $E_C \gg \tilde r$ and $J=1/(2N)$ for $E_C \ll \tilde r$. Geometrically, $\bd n$ lives on the hyperhoneycomb lattice of the NKSOS model and $\sigma\in\{1,-1\}$ indexes its two triangular sublattices. The distance between these lattices of minima is $\varphi^{\rm tot}(\sigma=1)-\varphi^{\rm tot}(\sigma=-1) = \pi \sqrt{2J}$ and is controlled by the ratio $E_C/\tilde r$. Plugging the minima's coordinates (\ref{eq:varphi_Ec_r},\ref{eq:varphi_c_Ec_r}) into the QBM action \eqref{eq:action_charge} and discretizing the  finally yields the NKSOS model for all $J$.

The general relations (\ref{eq:varphi_Ec_r},\ref{eq:varphi_c_Ec_r}) are also useful to understand the expression of the conductance in the NKSOS model. Within the QBM model, following Ref.~\cite{Paris_2026_sm} the (imaginary time) conductance in units of $e^2 K/h$ is expressed as
\begin{equation}
	G(i\omega)=\frac{2|\omega|}{K \pi(N-1)}\sum_a \langle |\phi^a(i\omega)|^2 \rangle.
\end{equation}
Going to the variables $\bd n$, $\sigma$, the previous expression is
\begin{equation}\label{eq:G_def}
	G(i\omega)=\frac{2\pi |\omega|}{K (N-1)}\left[ \langle |\bd n(i\omega)|^2 \rangle + J/2 \langle |\sigma(i\omega)|^2 \rangle\right],
\end{equation}
which is Eq.~(5) in the main text.

\section{III. Monte Carlo algorithm}
This section details the Monte Carlo algorithms used to study the NKSOS model. In order to drastically speed up the convergence time of the algorithm, we use cluster algorithms \cite{Swendsen_1987_sm,Wolff_cluster_algo_sm} in a form which is adapted to long range interactions \cite{Fukui_Todo_2009_sm,ClockFMetManonMichel_sm}. Cluster algorithms are reversible algorithms that crucially rely on the existence of an involution symmetry $R$ such that $R^2=1$ (e.g. the $\mathbb{Z}_2$ spin-flip symmetry in $O(n)$-spin models). For long range systems with $\beta$ degrees of freedom, a naive implementation of such an algorithm requires $O(\beta)$ operations to move a single degree of freedom (e.g. a spin in long range $O(n)$-spin models). However, it is known that this complexity can actually be reduced to $O(1)$ for most long-range interactions. The remainder of this section first discusses the implementation of a naive cluster algorithm for equivalent channels $r_a=r$, then presents its complexity reduction from $O(\beta)$ to $O(1)$, and performs an extensive performance analysis of the algorithm. Remarkably, the complexity-reduced cluster algorithm does not suffer from any critical slowing down, that is to say its run time is $O(\beta)$ which is the best theoretically achievable scaling. We also mention how the algorithm is modified for inequivalent channels $r_a \ne r$. Finally, we compare our Monte Carlo algorithm to established methods such as the numerical renormalization group.

\subsection{A. Integer-valued model}
The NKSOS model (Eqs.~(1,2) of the main text) is defined in terms of the fields $\sigma$ and $\bd n$, where $\bd n$ lives on the vertices of a hyperhoneycomb lattice. In practice, directly manipulating the coordinates $\bd n$ leads to small numerical round-off errors. To avoid this issue, we instead introduce integer-valued fields $\{M^a\}_{a\in \llbracket1,N\rrbracket}$ defined through
\begin{align}
	n_i^\alpha &= \sum_{a=1}^N G_a^\alpha M_i^a,\\
	\label{eq:sigma_M_def}
	\sigma_i &= 2\sum_{a=1}^N M_i^a + 1,
\end{align}
where $i$ denotes the imaginary-time index, $\alpha\in\llbracket 1,N-1\rrbracket$ labels the components of $\bd n$, and $\{\bd G_a\}_{a\in\llbracket 1,N\rrbracket}$ are the vectors introduced in Eq.~\eqref{eq:orthonormal}. In this representation, the allowed configurations $\bd M$ are such that $\sigma_i(\bd M_i)\in\{ -1,1\}$ and the NKSOS action with equivalent channels becomes
\begin{align}\label{eq:KSOS_M}
	S=\frac{1}{2}\sum_{i,j}S^{ij}(\bd M_i,\bd M_j),
\end{align}
where we have defined the pairwise interaction $S^{ij}(\bd M_i,\bd M_j)=S^{ji}(\bd M_j,\bd M_i)$ as
\begin{equation}
	S^{ij}(\bd M_i,\bd M_j)=C_{i-j}\left[\left(\bd M_i-\bd M_j\right)^2 -  \left(J-\frac{1}{2N}\right) \sigma_i\sigma_j\right]
\end{equation}
with $C_k=[K \left(\beta/\pi\right)^2 \sin^2\left(\pi k/\beta\right)]^{-1} + r(\delta_{k, 1}+ \delta_{k,- 1})$ for $k \ne 0$ and $C_0=0$.

\subsection{B. Cluster moves}
\label{sec:cluster_moves}
In order to build an efficient Monte Carlo algorithm, we implement cluster Monte-Carlo moves which perform collective updates involving many sites at once. We use the single-cluster variant of the algorithm, also known as the Wolff algorithm \cite{Wolff_cluster_algo_sm}, as it turned out to be faster than the multi-cluster one, also known as the Swendsen-Wang algorithm \cite{Swendsen_1987_sm}.

The single-cluster algorithm operates by creating a cluster, updating it as a single block with an operator $R$, and then repeating the process. For the algorithm to be correct, it is crucial that the operator $R$ be a symmetry of the pairwise interactions,
\begin{equation}\label{eq:S_ij_sym}
	S^{ij}(R(\bd M_i), R(\bd M_j))=S^{ij}(\bd M_i, \bd M_j),
\end{equation}
\begin{wrapfigure}{r}{0.3\textwidth}
	\centering
	\includegraphics[width=1\linewidth]{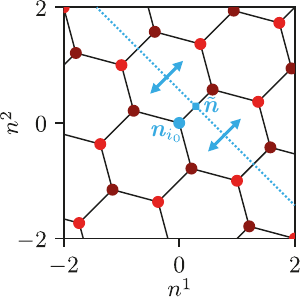}
	\caption{Involution $R^{\bd{\tilde M}}_a$ with $\bd {\tilde M}=(0,0,-1/2)$ and $a=3$ for the $N=3$ case. It consists in the reflection about the blue line and can be used to build a cluster rooted in $\bd n_{i_0}$.}
	\label{fig:cluster_involution}
	\vspace{-1.5cm}
\end{wrapfigure}
and that it be an involution, i.e. $R^2=1$. This construction is standard for Ising models, where spin clusters are built and subsequently flipped with the spin-flip operator $R(\sigma)=-\sigma$. Our algorithm follows the same philosophy, the spin-flip symmetry being replaced by a family of involutions acting on the $\bd M$ variables. For the NKSOS model, we consider the involutions $R_a^{\bd {\tilde M}}$ labeled by some reference coordinates $\bd  {\tilde M}$ and a channel $a\in\llbracket1,N\rrbracket$, and acting on $\bd M_j$ as
\begin{align}\label{eq:involution}
	R_a^{\bd {\tilde M}}(\bd M_j)&=-\bd M_j + 2\bd {\tilde M} + ((\bd M_j-\bd{\tilde M})\cdot \bd E_{bc})\bd E_{bc},
\end{align}
where $\bd E_{bc}=\bd e_b-\bd e_c$ with $\{\bd e_a\}_{a\in \llbracket 1,N\rrbracket}$ the canonical basis of $\mathbb R^N$, and $b\equiv a+1 \text{ mod } N$, $c\equiv a+2\text{ mod } N$. Strictly speaking, we consider Eq.~\eqref{eq:involution} only for $N>2$ and replace it by $R_a^{\bd {\tilde M}}(\bd M_j)=-\bd M_j + 2\bd {\tilde M}$ for $N=2$. For the following cluster algorithm, we will only be interested in half-integer coordinates \mbox{$\bd {\tilde M }\in (\mathbb{Z}/2)^N$} such that $\sigma(\bd{\tilde M})=0$. It is readily checked that for such points the involutions $R_a^{\bd {\tilde M}}$ satisfy Eq.~\eqref{eq:S_ij_sym} and $(R_a^{\bd {\tilde M}})^2=1$. In terms of $\bd n$ and $\sigma$, the involutions act as
\begin{align}\label{eq:involution_n_sigma}
	R_a^{\bd {\tilde M}}(\sigma_j)&=-\sigma_j,\\
	R_a^{\bd {\tilde M}}(\bd n_j)&=-\bd n_j + 2\bd {\tilde n} + ((\bd n_j-\bd{\tilde n})\cdot \bd G_{bc})\bd G_{bc},
\end{align}
with $\bd {\tilde n}=\sum_a \bd G_a {\tilde M}^a$, $\bd G_{bc}=\bd G_b - \bd G_c$ and $R_a^{\bd {\tilde M}}(\bd n_j)=-\bd n_j + 2\bd {\tilde n}$ for $N=2$. The first equation shows that $R_a^{\bd {\tilde M}}$ swaps the two sublattices defined by $\sigma=-1$ and $\sigma=1$, while the second is an operation on the hyperhoneycomb lattice which leaves unchanged the point $\bd{\tilde n}$. An example of an involution for $N=3$ is illustrated in Fig.~\ref{fig:cluster_involution}. In this case, $R_a^{\bd {\tilde M}}$ reduces to a reflection about the axis orthogonal to $\bd G_a$ and passing through ${\bd {\tilde n}}$. For larger $N$, the same construction generalizes to higher-dimensional rotations (for $N$ even) or reflections (for $N$ odd) which leave $\bd {\tilde n}$ unchanged.

The cluster algorithm then proceeds as follows.
\begin{enumerate}
	\item Pick uniformly a root site $i_0 \in \llbracket1,\beta\rrbracket$. Initialize both the stack of sites to check $S=\{ i_0\}$, and the cluster $C=\{i_0\}$. 
	
	\item Pick uniformly a label $a \in \llbracket 1, N \rrbracket$ and consider the involution $R=R^{\bd {\tilde M}}_a$ with $\bd {\tilde M}=\bd M_{i_0} - \sigma_{i_0} \bd e_a/2$. Notice that $\sigma(\bd {\tilde M})=0$ as anticipated.
	
	\item While $S$ is not empty:
	\begin{enumerate}
		\item Pop a site $i$ from the stack $S$.
		\item For every site $j \notin C$, add $j$ to $S$ and $C$ with probability
		\begin{align}\label{eq:p_act}
			P^{ij}_R(\bd M_i,\bd M_j)=1-\exp(-[S^{ij}(\bd M_i,R(\bd M_j))-S^{ij}(\bd M_i,\bd M_j)]_+).
		\end{align}
		where $[x]_+=\max(0,x)$.
	\end{enumerate}
	\item Once the stack $S$ is empty, all sites $i\in C$ belonging to the cluster are updated collectively according to $\bd M_i \to R(\bd M_i)$.
	
	\item Repeat from step 1.
\end{enumerate}
The correctness of the algorithm is guaranteed provided it satisfies the detailed balance condition and is irreducible, i.e. any configuration $\{\bd M_i\}_i$ can be reached from any other configuration $\{\bd M'_i\}_i$. The former condition can be checked using the standard procedure \cite{Wolff_cluster_algo_sm} which crucially relies on Eq.~\eqref{eq:S_ij_sym} and $(R_a^{\bd {\tilde M}})^2=1$. For the latter condition, we note that there is a non-zero probability that a cluster contains only its root $i_0$. In this case, using $\bd {\tilde M}=\bd M_{i_0} - \sigma_{i_0} \bd e_a/2$ the cluster move reduces to $\bd M_{i_0} \to R_a^{\bd {\tilde M}}(\bd M_{i_0})=\bd M_{i_0} - \sigma_{i_0} \bd e_a$. Such single-site updates clearly generate all possible configurations, thus proving irreducibility.

\subsection{C. Complexity reduction}
In the previous algorithm, a direct implementation of step 3.b consists in scanning all sites $j \notin C$. There are usually $\sim \beta$ such sites so step 3.b requires $O(\beta)$ operations. However, distant neighbors $j$ such that $|j-i|\gg 1$ are very rarely added to the cluster since $P^{ij}_R \sim S^{ij}\sim 1/|i-j|^2$. This intuitively shows that the naive implementation of step 3.b is far from optimal. We now detail a much more efficient implementation which only requires $O(\beta^0)=O(1)$ operations following Refs.~\cite{Fukui_Todo_2009_sm,ClockFMetManonMichel_sm}. The key idea is to add neighbors $j$ in two steps: first accepting them with a bound probability $\bar P^{ij}_R \ge P^{ij}_R$, and then accepting them with a resampling probability $P^{ij}_R/\bar P^{ij}_R$. To find the bound $\bar P^{ij}_R$, we explicitly write Eq.~\eqref{eq:p_act} for an involution $R=R_a^{\bd {\tilde M}}$ as $P_R^{ij}(\bd M_i,\bd M_j)=1-\exp[-\lambda^{ij}_R(\bd M_i,\bd M_j)]$ with
\begin{align}\label{eq:DeltaS}
	\lambda_R^{ij}(\bd M_i,\bd M_j)=C_{i-j} \left[ \left(2J - \frac{1}{N}\right) \sigma_i \sigma_j + \delta \bd M_i\cdot(\bd M_j - \bd {\tilde M})\right]_+
\end{align}
where $\delta M_i=4(\bd M_i - \bd{\tilde M}) - 2 (\bd M_i - \bd {\tilde M})\cdot\bd E_{bc} \bd E_{bc}$. An upper bound is $\bar P_R^{ij}=1-\exp[-\bar \lambda^{i-j}_R(\bd M_i,\bd M_+, \bd M_-)]$ with
\begin{equation}
	\bar \lambda^k_R(\bd M_i,\bd M_+, \bd M_-)=C_k\left[ \left|2J-\frac{1}{N}\right| + \sum_a \max(\delta M_i^a(M^a_+ - {\tilde M}^a),\delta M_i^a (M^a_- - {\tilde M}^a)) \right]_+,
\end{equation}
with bounds $M^a_+ \ge M_j^a$ and $M^a_- \le M_j^a$ for all $j$. Importantly, the bounds $\{\bar P^{ij}_R\}_j$ do not depend on the entire configuration $\{\bd M_j\}_j$ but only on $\bd M_i$, $\bd M_+$, $\bd M_-$ and $k=j-i$. Following Ref.~\cite{Fukui_Todo_2009_sm}, adding a neighbor $j=i+k$ with probability $\bar P^{ij}_R$ is then expressed as drawing an integer $n_k$ from the Poisson distribution $p(n_k,\bar \lambda_R^k) = e^{-\bar \lambda^k_R} (\bar \lambda^k_R )^{n_k} /n_k!$ of rate $\bar \lambda^k_R$ and adding site $j$ if $n_k>0$. Since the sum of independent Poisson variables remains Poisson distributed, the total number $n=\sum_{k=0}^{\beta-1} n_k$ follows the Poisson distribution $p(n,\bar \lambda_R)$ of rate $\bar \lambda_R = \sum_{k=0}^{\beta-1} \bar \lambda_R^k$. Each of the $n$ events generated is then assigned to a value $k$, and thus a neighbor $j=i+k$, using Walker's method of alias \cite{Walker1977_sm,Marsaglia2004gen_ran_var_sm} with probability $\bar \lambda^k_R/\bar \lambda_R=C_k/\sum_{l=0}^{\beta-1} C_l$. The optimized implementation of step 3.b thus proceeds as follows.
\begin{enumerate}
	\item Draw $n\in \mathbb{N}$ from the Poisson distribution of total rate $\bar \lambda_R$
	
	\item For $j=1,\dots,n$:
	\begin{enumerate}
		\item pick a site $j=i+k$ with probability $\bar \lambda^k_R/\bar \lambda_R=C_k/\sum_{l=0}^{\beta-1} C_l$ using Walker's method of alias.
		\item If $j \notin C$: add $j$ to the cluster $C$ and stack $S$ with probability $P^{ij}_R(\bd M_i,\bd M_j)/\bar P^{ij}_R(\bd M_i,\bd M_+, \bd M_-$).
	\end{enumerate}
\end{enumerate}
We now argue that this procedure indeed has $O(1)$ complexity. Since the following discussion is mainly technical, the next paragraph may be skipped on a first reading. First, one needs to compute
\begin{equation}
	\bar \lambda_R(\bd M_i,\bd M_+, \bd M_-)=\sum_{k=0}^{\beta-1} C_k \left[ \left|2J-\frac{1}{N}\right| + \sum_a \max(\delta M_i^a(M^a_+ - {\tilde M}^a),\delta M_i^a (M^a_- - {\tilde M}^a)) \right]_+.
\end{equation}
The sum $\sum_{k=0}^{\beta-1} C_k$ can be precomputed and is thus not an issue. Finding the bound $\bd M_+$ (respectively $\bd M_-$) can be done by setting $M_+^a=\max_j (M_j^a)$ (respectively $M_-^a=\min_j (M_j^a)$) every $\beta$ sites updated with $R$, and dynamically maintaining it as $M_+^a \to  \max(M_+^a,\max_{j\in C} (M_j^a))$ (resp. $M_-^a \to  \min(M_-^a,\min_{j\in C} (M_j^a))$) after each cluster move. This guarantees that the bounds stay relatively tight. Furthermore, since computing $\max_j (M_j^a)$ requires $\beta$ operations every $\beta$ operations and computing $\max_{j\in C} (M_j^a)$ requires $|C|$ operations every time a cluster of size $|C|$ is built, keeping track of $\bd M_+$, $\bd M_-$ does not spoil the $O(1)$ complexity. Once $\bar \lambda_R$ has been computed, one can draw $n \in \mathbb{N}$ and perform the "for" loop of step 2. The complexity of this loop is given by the average value of $n$, i.e. $\bar \lambda_R$. The sum $\sum_{k=0}^{\beta-1} C_k$ being finite in the limit $\beta \to \infty$, the complexity is $O(\bar \lambda_R) \sim O(\sum_a M^a_+ - M^a_-)$ which is numerically found to be $O(1)$ \footnote{The fact that $O\left(\sum_a (M^a_+ - M^a_-)\right)\sim O(1)$ has as much to do with the bounds being tight, i.e. $M^a_+ - M^a_- \sim \max_j (M^a_j) - \min_j (M^a_j)$, as with the physics of the system, i.e. $\max_j (M^a_j) - \min_j (M^a_j) \sim \beta^0 = 1$}. Finally, within the "for" loop the distribution $C_k/\sum_{l=0}^{\beta-1} C_l$ can be precomputed as it does not depend on the configuration $\{\bd M_i\}_i$ and therefore sampled in $O(1)$ time using Walker's method of alias, and computing $P^{ij}_R/\bar P^{ij}_R$ requires $O(1)$ operations. Thus, the overall complexity is $O(1)$.

The final complexity-reduced cluster algorithm (Clu-LR-KSOS) outputting $n_{\rm sample}$ samples and starting from an initial configuration $\{\bd M_i\}_i$ is summed up in Alg.~\ref{alg:NKSOS_MC}.

\begin{algorithm}
	\caption{Clu-LR-KSOS algorithm for equivalent channels}\label{alg:NKSOS_MC}
	{\bf Input} $\{ \bd M_i\}_i, n_{\rm sample}$\;
	${\rm Samples}=\{\}$\;
	$T_{\rm bounds}=0$\;
	\While{$n_{\rm sample}>0$}{
		\If{$T_{\rm bounds} \le 0$}{
			\For{$a=1,\dots,N$}{
				$M_+^a \gets \max_j (M_j^a)$\;
				$M_-^a \gets \min_j (M_j^a)$\;
				$T_{\rm bounds} \gets T_{\rm bounds} + \beta$\;
		}}
		$i_0={\rm choice}(\llbracket 1, \beta\rrbracket)$\;
		$S=\{i_0\}$\;
		$C=\{i_0\}$\;
		$a={\rm choice}(\llbracket 1, N\rrbracket)$\;
		$\bd {\tilde M} \gets \bd M_{i_0} - \sigma_{i_0} \bd e_a/2$\;
		$R \gets R_a^{\bd{\tilde M}}$\;
		\While{$S\neq \emptyset$}{
			$i\gets {\rm choice}(S)$\;
			$S \gets S \setminus \{ i\}$\;
			$n \sim p(n,\bar \lambda_R)$\;
			\For{$l =1,2,\dots, n$}{ Pick
				$k \in \llbracket 0, \beta-1\rrbracket$ according to $C_k/\sum_{l=0}^{\beta-1} C_l$ using a Walker table\;
				$j\gets  i+k$\;
				\If{$j \notin C$ and ${\rm ran}(0,1)< P_R^{ij}(\bd M_i,\bd M_j)/\bar P_R^{ij}(\bd M_i,\bd M_+, \bd M_-)$}{
					$C\gets C \cup \{j\}$\; $S\gets S \cup \{j\}$\; } }
		}
		\For{$i\in C$}{
			$\bd M_i\gets R(\bd M_i)$\;
			$T_{\rm bounds} \gets T_{\rm bounds} - 1$\;
			\For{$a=1,\dots,N$}{
				\uIf{$M_i^a > M^a_+$}{$M^a_+ \gets M_i^a$\;}
				\uElseIf{$M_i^a < M^a_-$}{      
					$M^a_- \gets M_i^a$\;}
		}}
		${\rm Samples}={\rm Samples}\cup \{\{\bd M_i\}_i\}$\;
		$n_{\rm sample}\gets n_{\rm sample} -1$}
	{\bf Return} ${\rm Samples}$\;
\end{algorithm}

\subsection{D. Performance test}
The performance of a Monte Carlo algorithm is reflected in the computational time needed to generate a new configuration $\{M_i\}_i$ which is independent of the previous ones. To capture the speedup achieved by the NKSOS model with the complexity-reduced cluster algorithm (Clu-LR-KSOS), we compare it to several other algorithms. For simplicity, all comparisons are performed at $K=1$ and in the Toulouse limit $J=0$, but results generalize to $K\ne 1$ and $J\ne 0$. The algorithms compared are:
\begin{itemize}
	\item Met-QBM : The Metropolis--Hastings algorithm for the Toulouse limit of the QBM model with $K=1$ and $\tilde r=10$ on a lattice
	\begin{align}
		S_{\rm BQM}=\frac{1}{2} \sum_{\substack{i,j=1 \\ i\ne j}}^\beta \frac{(\bd \varphi_i-\bd \varphi_j)^2}{\beta^2\sin^2\left(\frac{\pi(i-j)}{\beta}\right)} - \tilde r\sum_{a=1}^{N}\sum_{i=1}^\beta \cos \left(2\bd G_a\cdot \bd \varphi_i - \frac{\pi}{N}\right).
	\end{align}
	The algorithm picks uniformly a site $i$ and a transverse channel $\alpha$, proposes the move $\bd \varphi_i \to \bd \varphi'_i = \bd \varphi_i + \mathcal{N}(0,1)\bd e_\alpha$, and does the move with probability $P=\exp(-[S_{\rm QBM}(\bd \varphi') -S_{\rm QBM}(\bd \varphi)]_+)$.
	
	\item Met-KSOS : The Metropolis--Hastings algorithm for the NKSOS model \eqref{eq:KSOS_M} with $K=1$ and $r=0.2$. The value $r=0.2$ is such that the crossover scales $T^\star$ in the QBM and the KSOS are roughly equal, i.e. $T^\star_{\rm KSOS}(r=0.2)\simeq T^\star_{\rm QBM}(\tilde r=10)$, which ensures a fair comparison. The algorithm picks uniformly a site $i$ and a channel $a$, proposes the move $\bd M_i \to \bd M_i' = \bd M_i - \sigma_i \bd e_a$, and does the move with probability $P=\exp(-[S_{\rm KSOS}(\bd M') -S_{\rm KSOS}(\bd M)]_+)$.
	
	\item Clu-KSOS : The cluster algorithm for the NKSOS model with $K=1$ and $r=0.2$ as discussed in Section~\ref{sec:cluster_moves} but without the complexity reduction.
	
	\item Clu-LR-KSOS with $K=1$ and $r=0.2$.
\end{itemize}
\begin{figure}
	\includegraphics[width=0.48\textwidth]{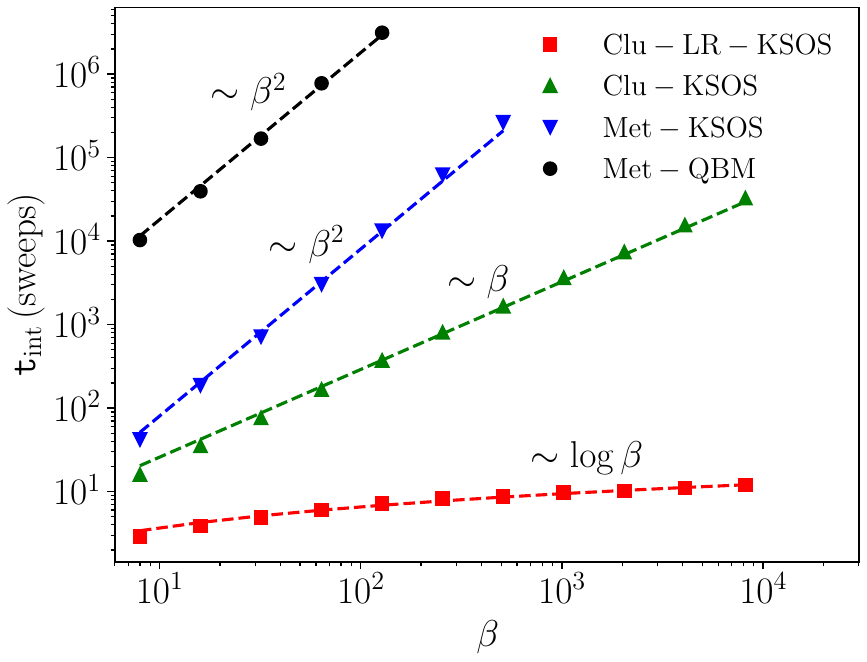}
	\includegraphics[width=0.48\textwidth]{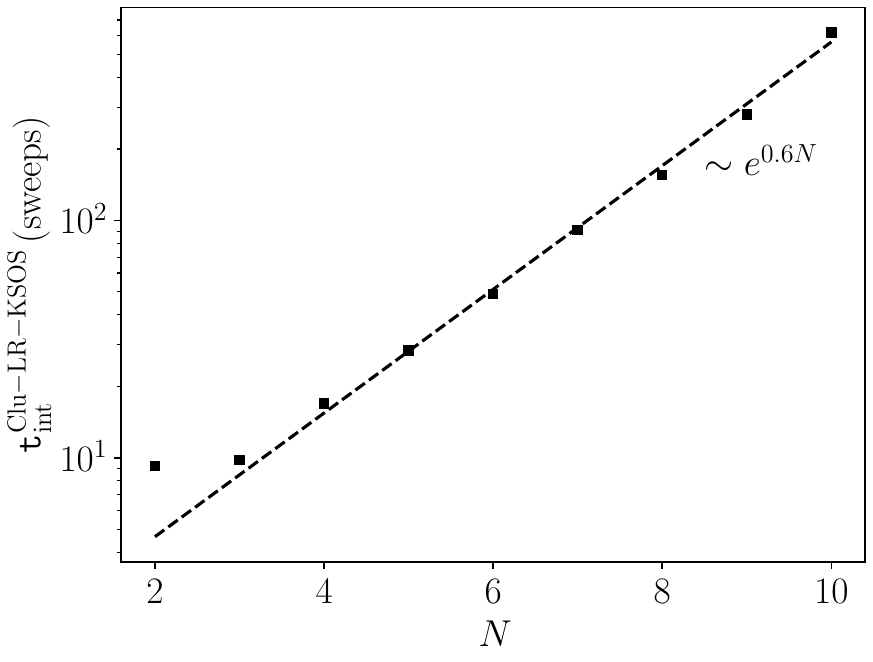}
	\caption{Left: Integrated auto-correlation time $\texttt t_{\rm int }$ for the 3CK at $J=0$, and $K=1$ for the Met-QBM at $\tilde r=10$ (black dots), the Met-KSOS (blue down triangles), the Clu-KSOS (green up triangles) and the Clu-LR-KSOS (red squares) algorithms at $r=0.2$. The dashed lines correspond to the scaling behaviors discussed in the main text. In particular, the red dashed line corresponds to a logarithmic scaling. Right: Integrated auto-correlation time $\texttt t_{\rm int }$ at $\beta=1024$ for the Clu-LR-KSOS algorithm as a function of the number of channels. The black dashed line is a guide to the eye.\label{fig:perf_test}}
\end{figure}
To test the performance of these algorithms, we focus on the variance of the lowest-frequency Fourier mode ${\cal O} = \langle |\bd \varphi(i2\pi/\beta)|^2 \rangle$ for the QBM and ${\cal O} = \langle |\bd n(i2\pi/\beta)|^2 \rangle$ for the NKSOS since it captures the large-scale physics which is responsible for slowing down Monte Carlo algorithms. We define an algorithmic time $\texttt{t}$ expressed in sweeps (i.e. $\beta$ operations) as increasing by $1/\beta$ each time a pairwise interaction $S^{ij}$ is explicitly evaluated. This leads to the definition of the autocorrelation function
\begin{equation}
	C_{\cal O}(\texttt{t})=\frac{\langle {\cal O}(\texttt{t}){\cal O}(0)\rangle - \langle {\cal O} \rangle^2}{\langle {\cal O}^2 \rangle - \langle {\cal O} \rangle^2},
\end{equation}
which generically decays exponentially. Its characteristic decay time is given by the integrated autocorrelation time $\texttt{t}_{\rm int}=1/2 \sum_{\texttt{t}=-\infty}^\infty C_{\cal O}(\texttt{t})$ which is the time needed to generate a new independent sample. Near a critical fixed point, $\texttt{t}_{\rm int} \sim \beta^z$ with $z \ge 0$ the (algorithmic) dynamical critical exponent. Critical slowing down is said to occur if $z>0$ and efficient algorithms are characterized by small values of $z$. The results for the four algorithms listed above are presented in Fig.~\ref{fig:perf_test}. Both the Met-QBM and the Met-KSOS have $z\simeq 2$, but the latter is faster by two orders of magnitude. More remarkably, the Clu-KSOS algorithm achieves $z\simeq 1$, while for the QBM model we were unable to obtain anything other than $z\simeq 2$ using analogous cluster algorithms. The cluster algorithms we tried rely on embedding the discrete variable $\bd n_i$ in the continuous one $\varphi_i$. The discrete variables are updated with cluster moves while the remaining fluctuations undergo Metropolis updates as in Refs.~\cite{Brower_1989_sm,Werner2005_sm}. The fact that the KSOS model lies in a different dynamical universality class than the QBM model while being in the same static universality class, i.e. both capture the Kondo physics, is the primary motivation for simulating the KSOS model over the QBM model. Finally, the complexity reduction yields $z\simeq 0$, thereby completely eliminating critical slowing down. Concretely, at $\beta=10^4$ our algorithm is $10^9$ times faster than a naive algorithm for the QBM. We also studied the dependency of the integrated autocorrelation time on the number of channels $N$ (Fig.~\ref{fig:perf_test}) and found that $\texttt t_{\rm int}$ increases exponentially with the number of channels. Nevertheless, the dynamical critical exponent remains $z \simeq 0$ whatever the value of $N$, making it possible to deal with a large number of channels.

Up until now we have focused on the case of equivalent channels $r_a=r$. For the case of nonequivalent channels, the operators $R_a^{\bd {\tilde M}}$ defined in Eq.~\eqref{eq:involution} are no longer symmetries of the action so the cluster algorithm breaks down. We thus replace it by a Metropolis--Hastings algorithm with a complexity reduction scheme similar to the one detailed above (see also Ref.~\cite{ClockFMetManonMichel_sm}). The resulting algorithm has $z\simeq 1$ since the complexity reduction scheme improves by $1$ the exponent of the Met-KSOS.

\subsection{E. Context and comparison with other methods}
In this section, we briefly discuss the place of our Monte Carlo approach among the methods commonly used to study quantum impurity problems, with particular emphasis on the multichannel Kondo model.

Among the most established numerical methods is the numerical renormalization group (NRG)~\cite{bulla2008numerical_sm}, which provides highly accurate access to both thermodynamic and dynamical properties and can accommodate general bath density-of-states structures. Its logarithmic discretization of the bath is particularly well suited to resolving physics over many decades of energy scales. In the context of the multichannel Kondo model, which is the focus of this work, NRG has proved successful and is commonly used to compare theoretical predictions with experimental results~\cite{iftikhar2018_sm,pouse2023quantum_sm}. However, its computational cost increases rapidly with the number of channels, making calculations at large $N$ challenging. Already for the 3CK model, substantial developments beyond the standard NRG scheme are required~\cite{mitchell2014generalized_sm}. Moreover, for the 3CK model, NRG calculations have so far been unable to access the high-transparency regime because of the proliferation of charge states on the metallic island~\cite{iftikhar2018_sm}.

Other approaches provide complementary information in more restricted regimes. Bethe-ansatz methods yield exact results for selected integrable Kondo models~\cite{andrei1984_sm,tsvelick1985exact_sm,affleck2001_sm}, while boundary conformal-field-theory methods provide exact information on universal properties at low-energy fixed points~\cite{affleck1991_sm}.

Monte Carlo methods, in contrast, provide a numerically unbiased treatment of equilibrium properties. They can directly access correlation functions in imaginary time. Real-frequency dynamical quantities can in principle be obtained by analytic continuation, although this introduces an additional numerical difficulty compared with NRG. Moreover, zero-temperature results generally require an extrapolation from finite-temperature calculations.

Regarding specifically the content of this work, the approach that is introduced provides a complementary framework tailored to a broad class of multichannel Kondo problems. Its main advantage is that the weak- and strong-tunneling regimes of charge-Kondo devices can be treated within a single representation. Our Monte Carlo algorithm also allows us to investigate substantially larger channel numbers than are currently accessible with multichannel NRG calculations, reaching up to $N=10$ in the present work. Furthermore, interactions in one-dimensional leads are incorporated naturally through the Luttinger parameter $K$, allowing us to study the evolution of multichannel Kondo physics for $K\neq1$. These features make the present approach particularly well suited to studying universal low-energy properties, crossover phenomena, and the dependence of multichannel Kondo physics on channel number and lead interactions.

\section{IV. Additional numerical results}

This section provides additional insights into the role of the anisotropy parameter $J$ and how to extract its scale-dependent renormalized value $J(j)$. We then discuss how to extract the zero-frequency conductance from its finite-size scaling. Finally, we compute the charge susceptibility of the 3CK model to demonstrate that the KSOS framework is not limited to the calculation of the conductance presented in the main text.

\subsection{A. Role of the anisotropy}

\begin{wrapfigure}{r}{0.35\textwidth}
	\centering
	\includegraphics[width=1\linewidth]{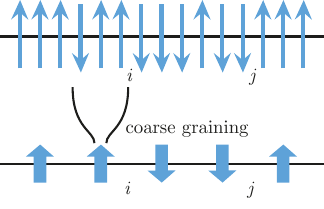}
	\caption{A microscopic spin configuration (top) has two distant instantons at $i$ and $j$ with small additional spin fluctuations. These fluctuations can be smeared out by coarse-graining the spins. This yields renormalized spins with a smaller length.}
	\label{fig:spin_RG}
\end{wrapfigure}

For $J>0$, we expect $J$ to be renormalized as we go to larger scales. We denote by $J(k)$ its renormalized value at a scale covering $k$ lattice sites. Since $J$ indexes the continuum of WT fixed points, we start by defining $J(k)$ at those fixed points. We consider the probability $\sim e^{-S}$ of a KSOS configuration having an instanton at site $i$ and an anti-instanton at site $j$. If we are right at the WT fixed point, fluctuations are totally suppressed and the field is totally flat apart from the jumps at $i$ and $j$. The spin part of the KSOS action then evaluates to
\begin{equation}
	e^{-S(\sigma)} \sim |i-j|^{-\frac{4J}{K}}.
\end{equation}
If we are not exactly at the WT fixed point, we still expect a power law behavior to emerge at large distances, albeit with a renormalized exponent $4J(k)/K$ at distances $k$. This power-low decay is what we consider as the very definition of $J(k)$ (for the Kondo action \eqref{eq:S_Kondo}, this is like defining $J(k)$ from the scaling dimension of the vertex operator $e^{i\gamma \vartheta^{\rm tot.}}$). To find the value of $J(k)$, we consider a typical field configuration with a distant instanton/anti-instanton pair at sites $i$ and $j$ as depicted in Fig.~\ref{fig:spin_RG}. We can get rid of the small spin fluctuations by coarse-graining the spin configuration. This comes at the cost of replacing the original spins $\sigma_i=\pm 1$ by renormalized ones with length $\sigma_i(k)=\pm m(k)$ where $m(k)$ is the magnetization at the scale $k$. It can be estimated as $m^2(k)=\langle (\sum_{i=1}^k \sigma_i)^2\rangle/k^2$ like in Ref.~\cite{Luijten_2001_sm} but we prefer the simpler alternative $m^2(k)=\langle \sigma_0 \sigma_k \rangle$. Since the renormalized model has no fluctuations below the scale $i-j$ one can directly compute that the probability of the instanton pair is
\begin{equation}
	e^{-S(\sigma)} \sim |i-j|^{-\frac{4J m^2(k)}{K}},
\end{equation}
which identifies
\begin{equation}\label{eq:J_j_def}
	J(k)=J m^2(k)=J \langle \sigma_0 \sigma_k \rangle.
\end{equation}

To obtain an RG flow diagram in the $(J,G)$ plane, we make use of Eqs.~(\ref{eq:G_def},\ref{eq:J_j_def}) which give the values of the conductance $G(i\omega)$ at the (frequency) scale $\omega$ and the anistropy $J(j)$ at the (time) scale $j$ \footnote{In the definition of $G(i\omega)$ in Eq.~\eqref{eq:G_def}, one must of course always use the microscopic anistropy $J$, and not the renormalized one $J(j)$.}. To match both scales, we associate the times $j=1,2,4,8,\dots,\beta/2$ to the frequencies $\omega=\pi,\pi/2,\pi/4,\pi/8,\dots,2\pi/\beta$ with $\beta=2^{17}$. The resulting flow diagram is displayed in Fig.~2(c) of the main text. It shows that the crossover from the WT phase to the NCK fixed point is not unique, but consists of a continuous family of curves parametrized by the anisotropy $J$. Therefore, collapsing Monte Carlo results at fixed $J\ne 0$ for different values of $r$ does not reconstruct a single RG trajectory but stitches together distinct RG trajectories. On the contrary, the line $J=0$ is stable under the RG, and the resulting crossover is universal. Results obtained following this procedure for the 2CK and the 3CK are displayed in Fig.~\ref{fig:crossovers}. Contrary to the weak-tunneling crossover, the crossover from the ST phase to the NCK fixed point is unique and, remarkably, independent of $J$, as shown by the comparison with the FRG results \cite{Paris_2026_sm}.

\begin{figure}[h!]
	\centering
	\includegraphics[width=0.48\linewidth]{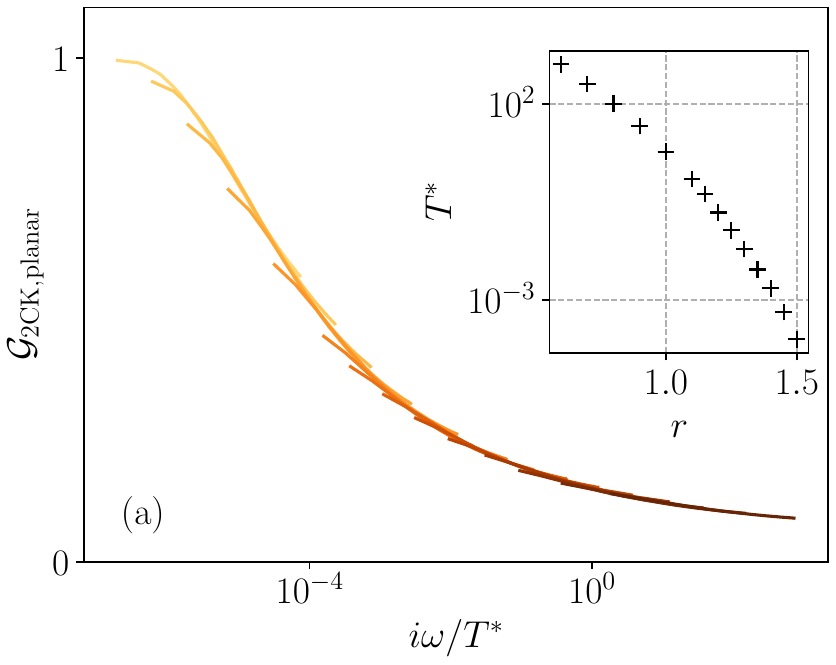}
	\includegraphics[width=0.48\linewidth]{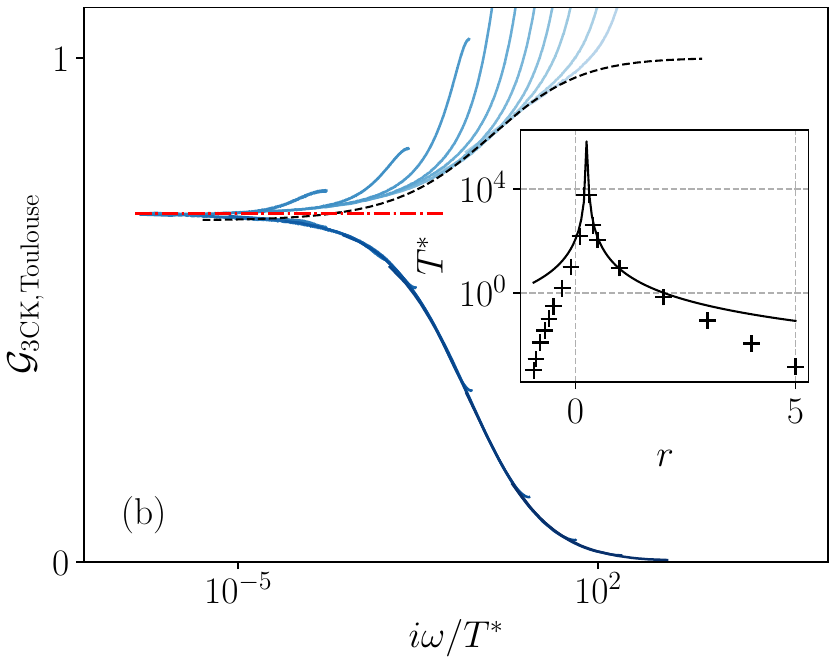}
	\\
	\includegraphics[width=0.48\linewidth]{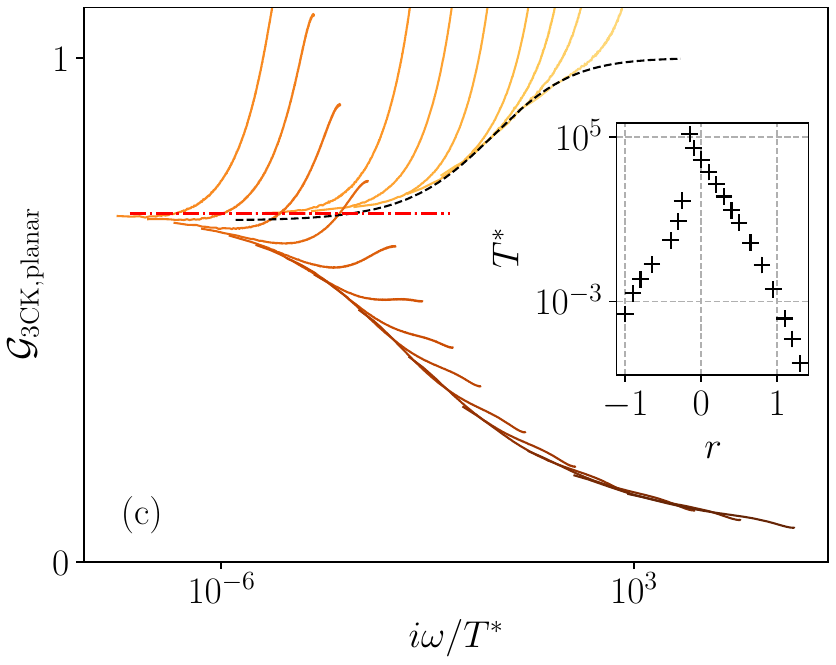}
	\includegraphics[width=0.48\linewidth]{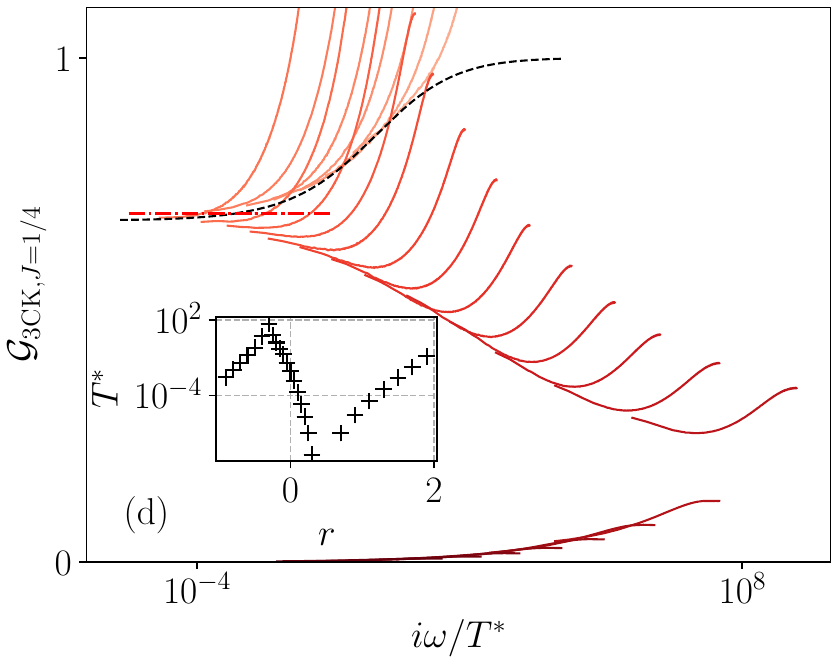}
	
	\caption{(a) Conductance of the planar 2CK ($J=1/4$) at $\beta=32768$, $K=1$ and various $r$. Finite size effects are still largely visible despite the large size. The high-frequency tails are partly erased for clarity. (b),(c),(d) Conductance of the 3CK at respectively $J=0,1/6,1/4$ at $\beta =8192$, $K=1$ and for various $r$. In the third case where $J=1/4>J_c=1/6$, a third conductance curve appears and goes to $0$ at low-frequencies since the WT phase is stable. FRG results are displayed with the dashed black line \cite{Paris_2026_sm}. The same curve is plotted in the three panels. Comparison with our numerical results show that the universal crossover from ST to NCK is independent from $J$. The red dashed and dotted line indicates the exact result at zero-frequency.}
	\label{fig:crossovers}
\end{figure}

\subsection{B. Finite-size scaling for the conductance}
Our Monte Carlo algorithm samples the Boltzmann distribution of the NKSOS model at a finite temperature $T=1/\beta$. The zero-temperature, zero-frequency conductance can be extracted from finite-size scaling using the known asymptotic behavior~\cite{affleck2001_sm,Yi_1998_sm,Yi_2002_sm}
\begin{equation}\label{eq:scaling}
	G(i\omega_1=i2\pi T,T)=G^*+aT^\nu,
\end{equation}
where $\omega_1=2\pi T$ is the first nonzero Matsubara frequency, $a\in\mathbb{R}$, $G^*\in[0,1]$, and $\nu>0$. The parameters $a$, $G^*$, and $\nu$ are obtained from a three-parameter fit of the finite-temperature data. To reduce finite-size effects, the smallest system sizes (corresponding to the highest temperatures) are excluded from the fit. This procedure is used, for example, to obtain Fig.~2(d) of the main text. Since it is not evident that the finite-temperature data converge towards the indicated zero-temperature conductance values, we replot the data as a function of $(\beta/32)^{-\nu}$ in Fig.~\ref{fig:G_K}. This parametrization makes the convergence more transparent, since Eq.~\eqref{eq:scaling} predicts a linear dependence on this variable for a given $N$. The exponent $\nu$, which depends on the number of channels, is determined numerically from the fit to Eq.~\eqref{eq:scaling}. The excellent agreement between the fitting functions and the raw data provides strong evidence for the validity of the finite-temperature scaling procedure.

\begin{figure}
	\centering
	\includegraphics[width=0.5\linewidth]{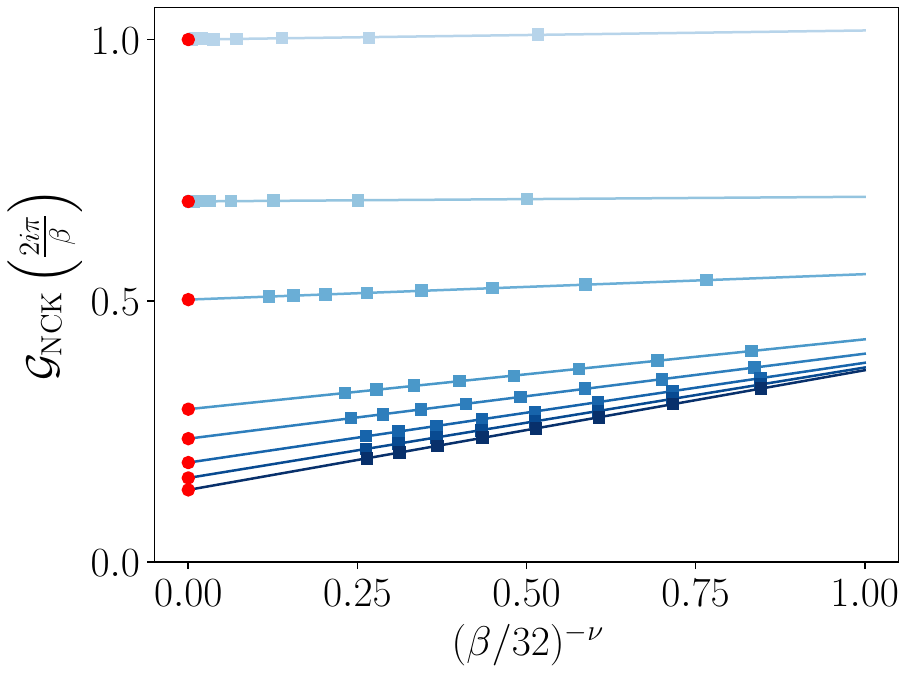}
	\caption{From top to bottom, the conductance of the 2KSOS to 10KSOS models as a function of $(\beta/32)^{-\nu}$ with $\beta=64,128,256,\dots$, where $\nu$ depends on the number of channels. Square markers denote the raw finite-temperature Monte Carlo data; error bars are smaller than the markers. The solid lines show the scaling form~\eqref{eq:scaling}, with $G^*$, $\nu$, and $a$ determined from a three-parameter fit using data for $\beta\geq64$. The fit describes the raw data excellently. The corresponding zero-temperature extrapolations $G^*$ are shown as red circles.}
	\label{fig:scaling}
\end{figure}

Determining the full conductance curve $G^*(K)$ shown in Fig.~3 of the main text requires a slightly more elaborate extrapolation scheme. For convenience, the KSOS action is rewritten as
\begin{align}\label{eq:action_KSOS}
	S_0=&\frac{1}{2K}\sum_{\substack{i,j=1\\ (i\ne j,i\ne j\pm1)}}^\beta\frac{\left(\bd n_i-\bd n_j\right)^2 - J \sigma_i\sigma_j }{\left(\frac{\beta}{\pi}\right)^2 \sin^2\left(\frac{\pi(i-j)}{\beta}\right)} ,\\
	\label{eq:action_S_r}
	S_r=& r_{\rm MC}\sum_{i=1}^\beta \left[\left(\bd n_i-\bd n_{i+1}\right)^2 - J \sigma_i\sigma_{i+1}\right].
\end{align}
where the parameter $r_{\rm MC}$ contains both the jump cost $r$ and the nearest-neighbor contribution of the imaginary-time kernel,
\begin{align}
	r_{\rm MC}=r+\frac{1}{K}\frac{\pi^2 }{\beta^2 \sin^2\left(\frac{\pi}{\beta}\right)}.
\end{align}
A simple strategy would be to fix a single value of $r_{\rm MC}$, independent of $K$, perform simulations at finite temperatures $\{T_m\}_m$, and then extrapolate the conductances $\{G(i2\pi T_m,T_m)\}_m$ to $T\to0$. In practice, this procedure works well when the conductances evolve monotonically with temperature. To meet this criterion, we divide the $K$ axis into two overlapping regions and use a different value of $r_{\rm MC}$ in each region (see Fig.~\ref{fig:G_K}). Independent zero-temperature extrapolations are then performed in each region. The final conductance curve $G^*(K)$ is obtained by combining the two extrapolations and performing a weighted average in the overlap region. The resulting conductance curve is shown in Fig.~\ref{fig:G_K}. As above, it is not obvious from Fig.~\ref{fig:G_K} that the finite-temperature conductance converges towards its asymptotic value. Hence, the fit of the raw data by the scaling function~\eqref{eq:scaling} is displayed in Fig.~\ref{fig:G_K}. The excellent agreement between the fitting functions and the finite-temperature data ensures the validity of the line of asymptotic value (red-dashed line) shown in Fig.~\ref{fig:G_K}.

\begin{figure}[h]
	\centering
	\includegraphics[width=0.49\linewidth]{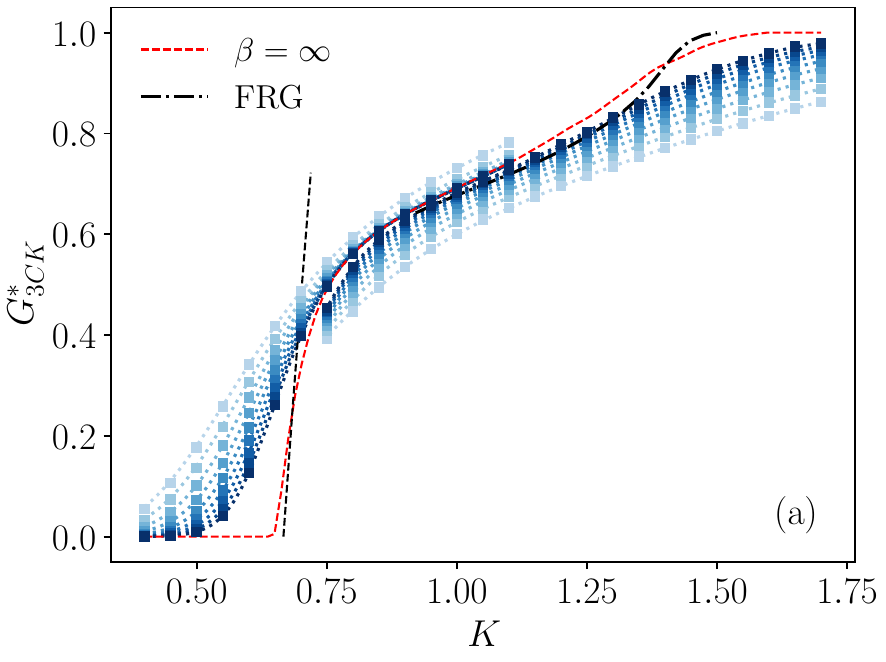}
	\includegraphics[width=0.48\linewidth]{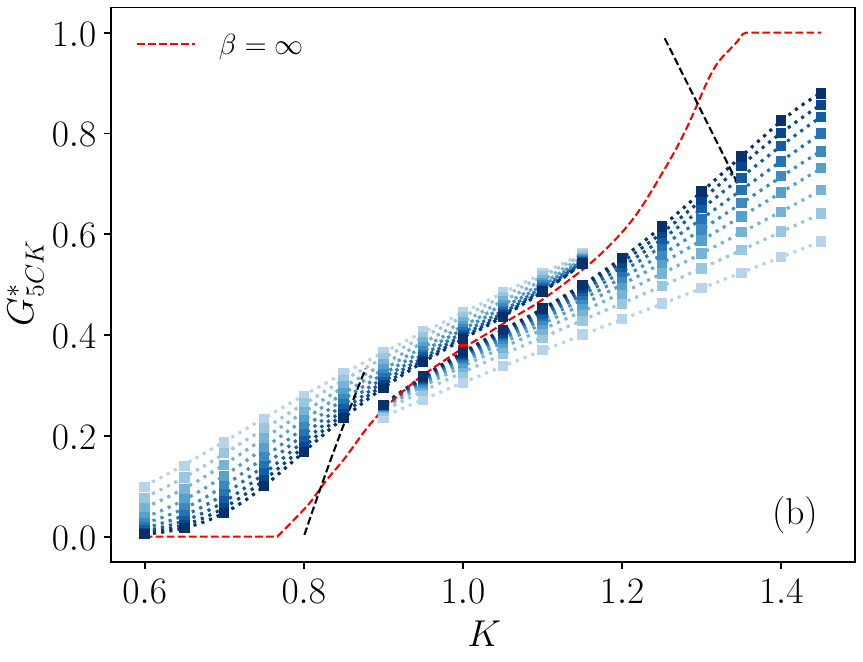}\\
	
	\vspace{3mm}
	\includegraphics[width=0.24\linewidth]{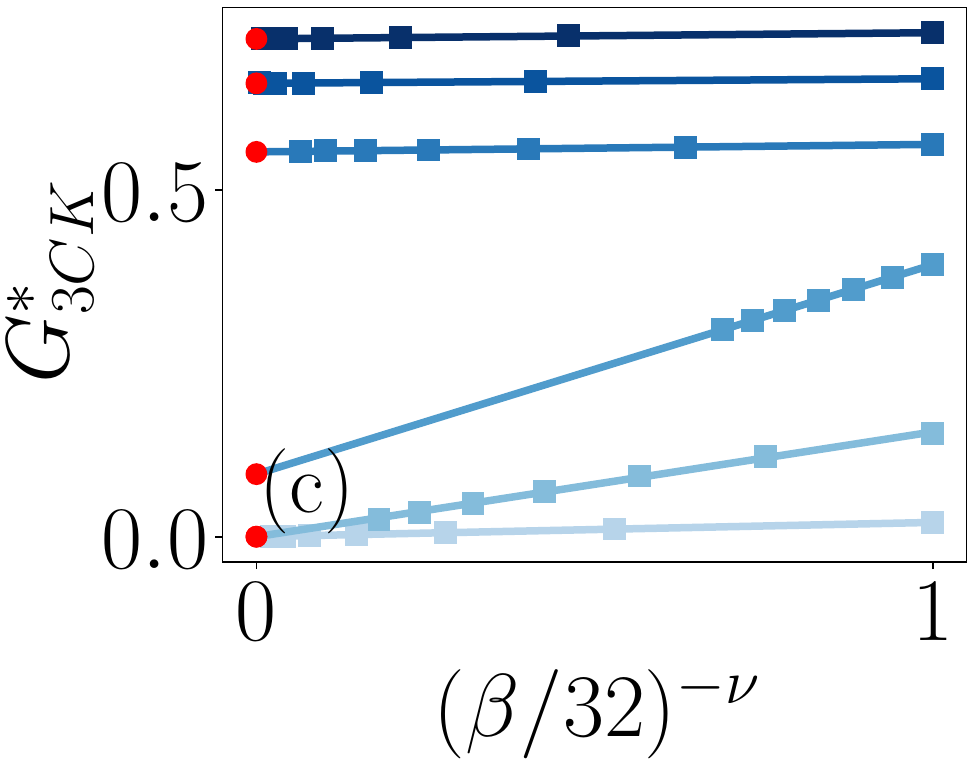}
	\includegraphics[width=0.24\linewidth]{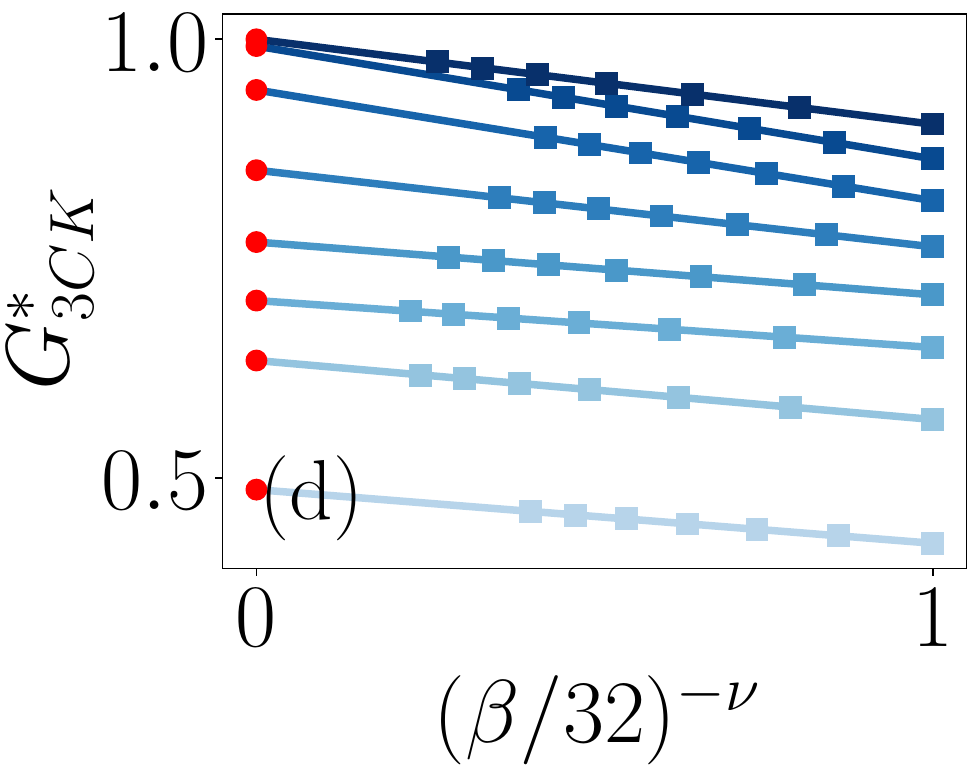}
	\includegraphics[width=0.24\linewidth]{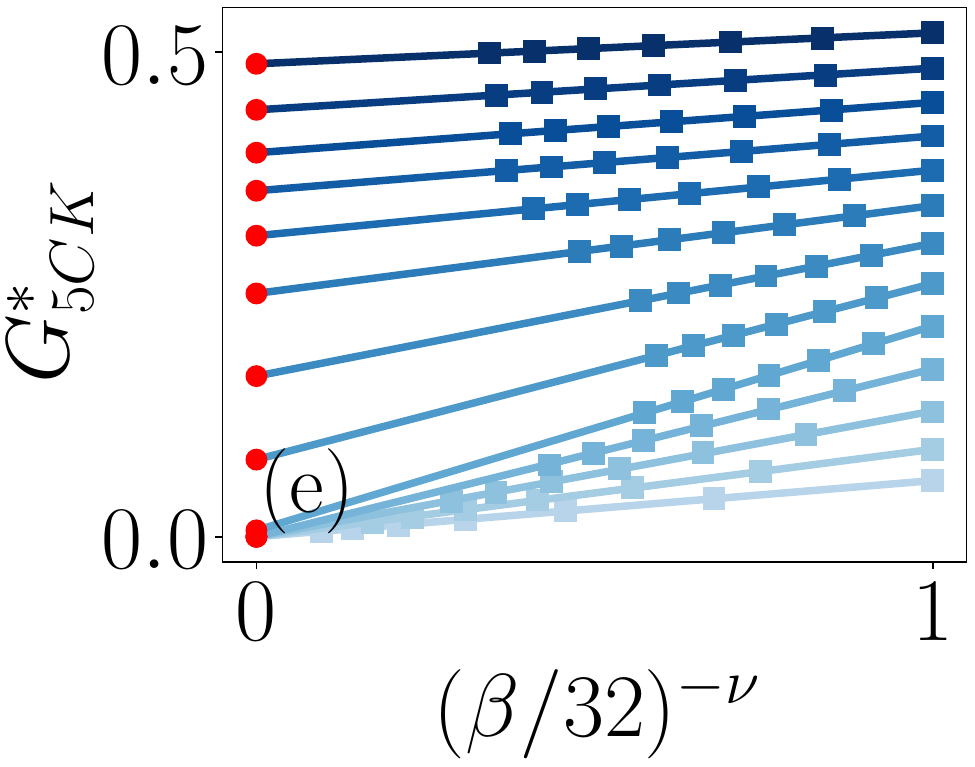}
	\includegraphics[width=0.24\linewidth]{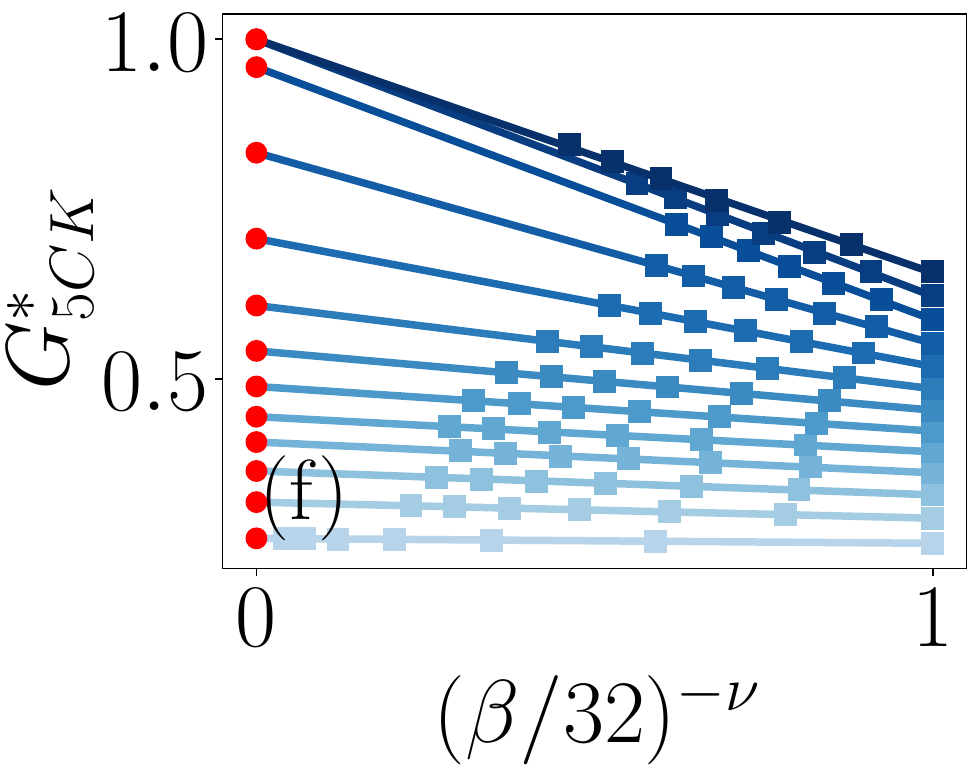}
	\caption{(a): Conductance obtained from the NKSOS as a function of $K$ for the 3CK. Blue symbols correspond to the data obtained with the Monte Carlo simulation at $\beta\in\{8,\dots,2048\}$ at $r_{\rm MC}=1.2$ (left set of curves) and $r_{\rm MC}=1.5$ (right set of curves). These data are processed with the multi-histogram method (MHM) to obtain the dotted blue lines, that are continuous in $K$. MHM curves are then extrapolated in the limit $\beta\to \infty$ (in red). The dashed black line represents perturbative RG results. The FRG results of  Ref.~\cite{Paris_2026_sm} are displayed with the dashed and dotted black line. (b): Same figure for the 5CK model. The left set of curves is obtained at $r_{\rm MC}=1.15$, the right one at $r_{\rm MC}=1.45$. (c): Conductance of the 3CK model with interacting leads for uniformly spaced values of $K\in[0.4,1.1]$, corresponding to the left set of blue curves in Fig.~(a). The blue lines show the corresponding fits to Eq.~\eqref{eq:scaling}, with the exponent $\nu$ determined independently for each $K$. (d): Same as (c), for $K\in[0.75,1.7]$, corresponding to the right set of curves in Fig.~(a). (e): Conductance of the 5CK model for uniformly spaced values of $K\in[0.6,1.15]$, corresponding to the left set of curves in Fig.~(b). (f): Same as (e), for $K\in[0.9,1.45]$, corresponding to the right set of curves in Fig.~(b).}
	\label{fig:G_K}
\end{figure}

\subsection{C. Determination of the universal conductance curve}
The conductance depends on the microscopic parameters of the KSOS model, i.e. the values of $r$ and of the lattice spacing (taken equal to $1$). Nevertheless, it exhibits a universal scaling behavior in the low-frequency limit $|\omega|\ll 1$, as a function of $i\omega/T^*$, where $T^*$ is a function of the microscopic parameters, as shown in Eq.~(6) of the main text.

In order to determine the full function $\mathcal{G}(i\omega/T^*)$, one can either vary $i\omega$ or play with $T^*$ by tuning the microscopic parameters $r$ of the KSOS model. In the KSOS model, the frequency $\omega$ is constrained to $2\pi/\beta<|\omega|<\pi$. Since the simulations are performed at finite (though small) temperatures, we find it convenient to compute the conductance for different values of $r$. The corresponding values of $T^*$ are then determined by collapsing all the curves, as illustrated in Fig.~\ref{fig:collapse}.

The scaling form of $G(i\omega)$ is not expected to hold at frequencies of the order of the ultraviolet cutoff. Therefore, the upward bends of the curves in Fig.~\ref{fig:collapse} come from the finite lattice spacing, that is to say it is a finite-cutoff effect. They are non-universal as they are not predicted by the scaling form Eq.~(6) of the main text and their onset in terms of $r$ is generally not known. Such effects are also expected in the $N$-channel Kondo model or quantum Brownian motion model and might differ from those of the KSOS model as they are, by definition, not universal. However, below this non-universal high-frequency regime, all models should display the same universal curve as they are in the same universality class.

\begin{figure}
	\centering
	\includegraphics[width=0.48\linewidth]{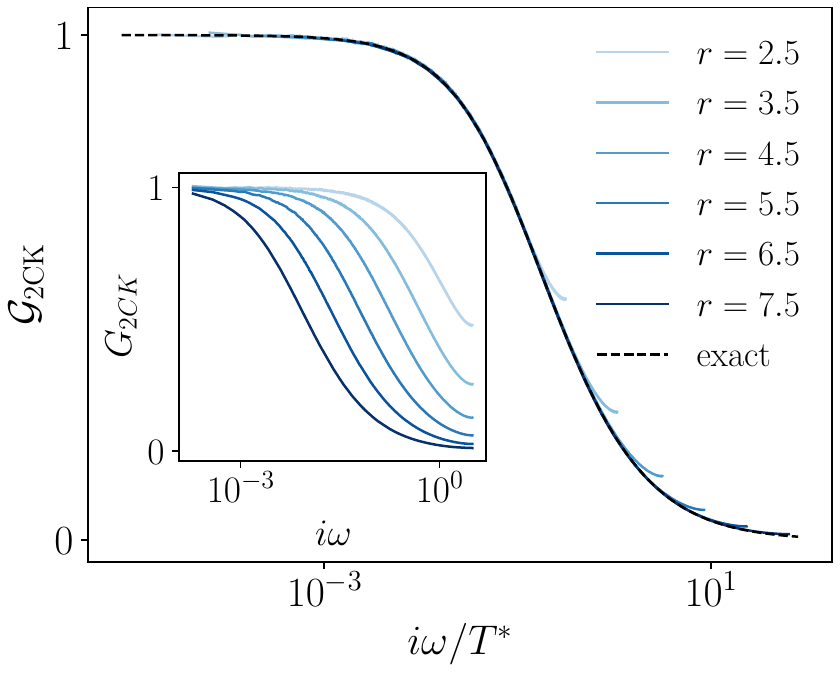}
	\includegraphics[width=0.48\linewidth]{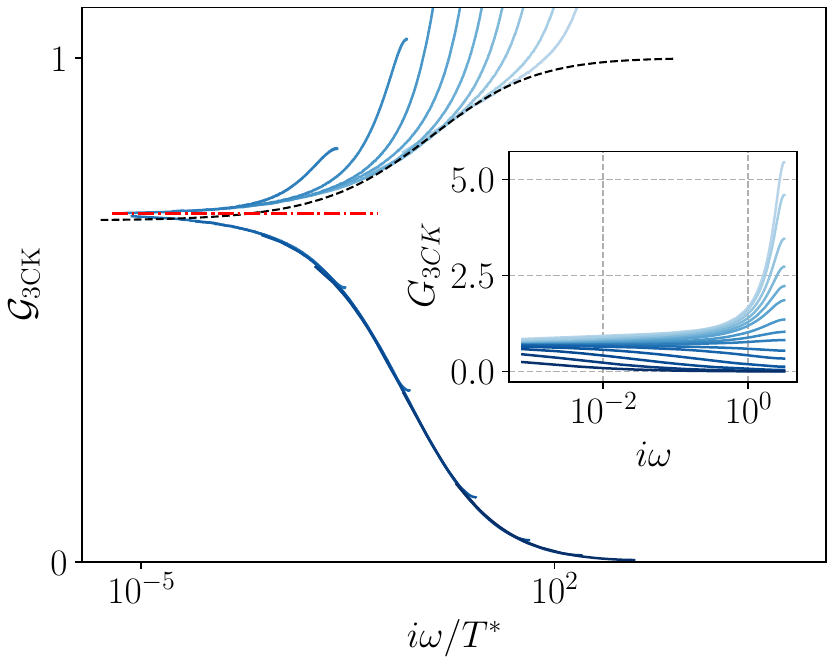}
	\caption{Left: Conductance of the 2KSOS model as a function of $i\omega/T^*$. The inset displays the same conductance data but as a function of $i\omega$, i.e. without the rescaling by the crossover energy scale $T^*$. The corresponding values of $T^*$ are shown in Fig.~2 of the main text. Right: Same figure but for the 3KSOS model for $r\in\{1.1,0.9,0.7,0.5,0.4,0.3,0.2,0.1,0.05\}$ for the top curve and $r\in\{1.5,2,3,4,5,6\}$ for the bottom curve. The dashed black line indicated the FRG results in the strong tunneling regime.}
	\label{fig:collapse}
\end{figure}

\subsection{D. Computation of the charge susceptibility}

Our Monte Carlo algorithm can evaluate the expectation value $\langle \mathcal{O} \rangle$ of any thermodynamic observable $\mathcal{O}$ which can be expressed in terms of the variables $\sigma$ and $\bd n$. Since the $N$-channel Kondo model and charge Kondo model map onto the NKSOS, this can be extended to any thermodynamic observables expressed in these two other models.

As a concrete example, we compute here the charge susceptibility of the charge Kondo model which is the analogue of the magnetic susceptibility of the spin Kondo model ~\cite{piquard2023observing_sm}. Up to an overall prefactor, it is given by
\begin{equation}
	\chi=\frac{T}{E_C^2}\left.\frac{\partial^2 \ln Z}{\partial N_g^2}\right|_{N_g=1/2}
\end{equation}
where $Z$ is the partition function of the charge Kondo model Eq.~\eqref{eq:charge_Kondo} with the generalized charging energy $E_c(\hat{N}-N_g)^2$. To express this observable in terms of the KSOS variables $(\bd n,\sigma)$, we follow the mapping from the charge Kondo model to the QBM model and finally to the NKSOS model outlined in Sec.~\ref{sec:QBM}. In terms of the QBM action Eq.~\eqref{eq:action_charge} with generalized charging energy $E_C\int d\tau (\sum_a \phi^a/\pi + N_g)^2$, the susceptibility is explicitly given by
\begin{equation}
	\chi=\frac{2}{E_C} + 4 T \int_0^\beta d\tau \int_0^\beta d\tau' \left\langle \left(\frac{1}{\pi} \sum_a \phi^a(\tau) +\frac12\right)\left(\frac{1}{\pi} \sum_a \phi^a(\tau') +\frac12\right)\right\rangle_c,
\end{equation}
where $\langle \cdot \rangle_c$ denotes a connected correlation function. To go to the NKSOS model, we discretize imaginary time as $\int d\tau \to \sum_i$ and the fields as $\phi^a(\tau) \to \pi M_i^a$. This leads to
\begin{equation}
	\chi= \frac{2}{E_C} + 4 T \sum_{i,j=1}^\beta \left\langle \left(\sum_a M^a_i +\frac12\right)\left(\sum_a M^a_j +\frac12\right)\right\rangle_c.
\end{equation}
Finally, using Eq.~\eqref{eq:sigma_M_def}, one arrives at
\begin{equation}\label{eq:susceptibility_KSOS}
	\chi= \beta \left[\langle m^2 \rangle -\langle m \rangle^2 \right] \text{ with } m=\frac{1}{\beta} \sum_{i=1}^\beta \sigma_i
\end{equation}
where we have introduced the magnetization $m$ and dropped the constant term since it is not responsible for the divergence of $\chi$. Equation~\eqref{eq:susceptibility_KSOS} takes the exact same form as the susceptibility of a classical Ising model. The only difference is that, in the KSOS model, $\beta$ plays both the role of the inverse temperature and the system size.

Near the NCK fixed point, we expect the susceptibility $\chi$ to diverge in the zero-temperature limit. For instance, at $K=1$, $\chi\sim \ln \beta$ in the 2CK, while $\chi \sim \beta^{1/5}$ in the 3CK~\cite{Vojta2006_sm}. For the 3CK, Fig.~\ref{fig:chi_beta} reproduces this result. By fitting the lowest-temperature data, we recover the exponent $1/5=0.2$ with a 1.5\% error.

\begin{figure}
	\centering
	\includegraphics[width=0.5\linewidth]{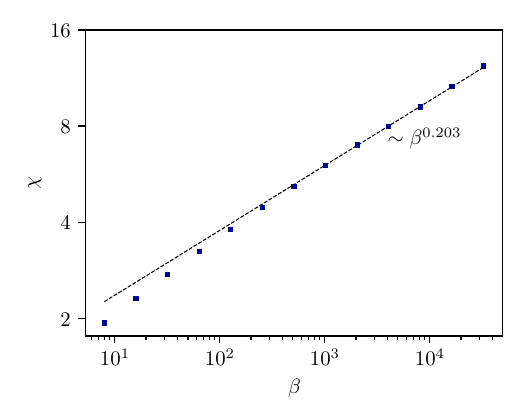}
	\caption{Charge susceptibility $\chi$ as defined in Eq.~\eqref{eq:susceptibility_KSOS} for the 3KSOS, at $K=1$, $r=0$, $J=0$ and $\beta=8,16,32,\dots,32768$. The dashed line is a power-law fit over the 5 smallest temperatures. Error bars are shown but smaller than the marker size.}
	\label{fig:chi_beta}
\end{figure}

\section{V. Emery--Kivelson solution of the two-channel Kondo model\label{app:EK}}
In this section, we derive an exact expression for the conductance of the 2CK model at the Toulouse point $\lambda^z=2\pi$ following Emery--Kivelson's refermionization \cite{Emery_1992_sm}.

\subsection{A. Mapping on the boundary sine-Gordon model}
The 2CK model in Eqs.~(\ref{eq:S_Kondo},\ref{eq:S_Kondo2}) can be rewritten in the Toulouse limit $\gamma=0$ as
\begin{equation}
	S_{\rm 2CK}=\frac{K}{2\pi} \sum_{\omega_n} |\omega_n| |\vartheta(i\omega_n)|^2 - 2t \sigma_x \int_{0}^{\beta} d\tau  \cos \vartheta(\tau),
\end{equation}
where we have dropped the total mode $\vartheta^{\rm tot}$ which decouples from $\vartheta$ and rescaled $\vartheta(\tau) \to \vartheta(\tau)/|\bd G_1 |= \sqrt{2}\vartheta(\tau)$. The partition function of the model is then computed in the eigenbasis of $\sigma_x$ as
\begin{align}
	Z_{\rm 2CK}&={\rm Tr}_\sigma \int \mathcal D\vartheta\, e^{-S_{\rm 2CK}}= 2 \int \mathcal D\vartheta\, e^{-S_{\rm bsG}}=2Z_{\rm bsG},
\end{align}
where $Z_{\rm bsG}$ is the partition function of the boundary sine-Gordon model
\begin{equation}\label{eq:bsG}
	S_{\rm bsG}=\frac{K}{2\pi}\sum_{\omega_n}|\omega_n||\vartheta(i\omega_n)|^2+2t\int_{0}^{\beta} d\tau\cos\vartheta(\tau).
\end{equation}
The correlation functions of the 2CK can thus be computed via the boundary sine-Gordon model. 

\subsection{B. A convenient identity for the conductance}
The conductance of the 2CK in the Toulouse limit is given by
\begin{equation}
	G(i\omega,T=0)=\frac{2\pi |\omega|}{K}\langle n(i\omega) n(-i\omega)\rangle =1-\frac{K|\omega|}{\pi}\langle \vartheta(i\omega)\vartheta(-i\omega)\rangle.
\end{equation}
The last equality is given in Ref.~\cite{Yi_2002_sm} and can be derived via Coulomb gas and instanton expansions much like the mapping between the KSOS and Kondo models. The correlation function $\langle|\vartheta(i\omega)|^2\rangle$, and thus the conductance $G(i\omega,T=0)$, can be computed using an identity we now derive. Adding an external source to the partition function, we find
\begin{align}
	Z[h]&=\int \mathcal D\vartheta \exp\left[-S+\frac{K}{2\pi}\int \frac{d\omega}{2\pi}|\omega|\left(\vartheta(i\omega)h(-i\omega)+{\rm H.c.}\right)\right]\label{eq:Z1}\\
	&=\int \mathcal D\vartheta \exp\left[-\frac{K}{2\pi}\int\frac{d\omega}{2\pi}|\omega|\left(|\vartheta(i\omega)|^2-|h(i\omega)|^2\right)-2t\int_{-\infty}^{\infty} d\tau \cos(\vartheta+h)\right].\label{eq:Z2}
\end{align}
Computing $\frac{\delta^2\ln Z}{\delta h(i\omega)\delta h(-i\omega)}$ using Eqs.~(\ref{eq:Z1},\ref{eq:Z2}), we obtain the identity:
\begin{align}
	G(i\omega,T=0)=1-\frac{K|\omega|}{\pi}\langle \vartheta(i\omega)\vartheta(-i\omega)\rangle
	&=- \frac{2\pi t}{K|\omega|}\left\langle\cos(\vartheta(0))\right\rangle - \frac{4\pi t^2}{K|\omega|}\int_{-\infty}^{+\infty} d\tau\, e^{i\omega\tau}\left\langle\sin\vartheta(\tau)\sin\vartheta(0)\right\rangle.
\end{align}

\subsection{C. Refermionization}
At $K=1$, the scaling dimension of the cosine is exactly $1/2$. This indicates that the model can be refermionized. The Hamiltonian associated with the action~\eqref{eq:bsG} reads
\begin{equation}
	H=\frac{1}{4\pi}\int_{-\infty}^{+\infty} dx (\partial_x\hvartheta)^2+2t\cos\hvartheta(x=0),
\end{equation}
with $[\hvartheta(x),\hvartheta(x')]=-i\pi{\rm sign}(x-x')$. Following Ref.~\cite{furusaki_1995_sm}, it can be refermionized via the correspondence 
\begin{equation}
	e^{i\hvartheta(x)}=\sqrt{2\pi\alpha}\,\hat\eta\,\hpsi(x),
\end{equation}
where $\hat\eta$ is a Majorana operator with $\hat\eta^2=1$ and $\hpsi$ a chiral fermion satisfying $\{\hat\eta,\hpsi^{(\dagger)}(x)\}=0$. Writing $\hpsi(x)=\frac{1}{\sqrt{2}}(\hat\xi_1(x)+i\hat\xi_2(x))$ with $\hat\xi_1$, $\hat\xi_2$ Majorana fermions, we obtain the quadratic Hamiltonian
\begin{align}
	H=\frac{iv_F}{2}\int dx \left[\hat\xi_1\partial_x\hat\xi_1+\hat\xi_2\partial_x\hat\xi_2\right]+i\sqrt{2}\tilde\lambda\hat\eta\hat\xi_2(0),
\end{align}
with $\tilde \lambda=t\sqrt{2\pi \alpha}$. The bosonic correlation functions can then be deduced from the fermionic propagators:
\begin{align}
	\frac{2\pi t}{|\omega|}\langle \cos\vartheta(0)\rangle=i\sqrt{\pi\alpha}\frac{2\pi t}{|\omega|}\langle \eta(0)&\xi_2(x=0,0)\rangle=-\frac{\tilde\lambda^2}{2|\omega|}\int_{-\infty}^{+\infty}d\omega\frac{1}{|\omega|+\tilde \lambda^2},\\
	\frac{4\pi t^2}{|\omega|}\int_0^\infty d\tau \, e^{i\omega\tau}\left\langle\sin\vartheta(\tau)\sin\vartheta(0)\right\rangle&=\frac{2\pi \tilde \lambda^2}{|\omega|} \int_{-\infty}^{+\infty}\frac{d\omega'}{2\pi}\langle \eta(i\omega')\eta(-i\omega')\rangle \langle \xi_1(x=0,i\omega-i\omega')  \xi_1(x=0,i\omega'-i\omega)\rangle\nonumber\\
	&=-\frac{\tilde \lambda^2}{|\omega|}\ln\left(\frac{|\omega|+\tilde \lambda^2}{\tilde \lambda^2}\right) + \frac{\tilde \lambda^2}{2|\omega|} \int_{-\infty}^{+\infty}d\omega\frac{1}{|\omega|+\tilde \lambda^2},
\end{align}
where we took the limit $\beta \to \infty$.
Putting all the pieces together, the conductance is given by
\begin{align}
	G(i\omega,T=0)&=\frac{\tilde\lambda^2}{|\omega|}\ln\left(\frac{|\omega|+\tilde \lambda^2}{\tilde \lambda^2}\right).
\end{align}
The Kondo temperature is therefore $T^*=\tilde \lambda^2=2\pi\alpha t^2=2\pi\alpha e^{-r}$.

\section{VI. $N$-channel Kondo model in the limit of large $N$}
In this section, we map explicitly the $N$-channel Kondo model in the $N\to \infty$ limit onto $N$ decoupled boundary sine-Gordon models. This is achieved by starting from the partition function associated to the action~\eqref{eq:action_charge}. One first isolates the total mode $\lambda=\sum_a \phi_a$ as
\begin{align}
	Z&=\int \mathcal D\bd \phi \mathcal D \lambda \exp\left[-\sum_{a=1}^{N}S_{\rm bsG}[\phi^a] - \int_0^\beta d \tau E_C\left(\frac{\lambda(\tau)}{\pi} + \frac12\right)^2 \right]\delta\left(\lambda - \sum_{a=1}^N\phi^a\right),
\end{align}
where $S_{\rm bsG}[\phi]=\frac{1}{\pi K}\sum_{\omega_n}|\omega_n|| \phi^a(i\omega_n)|^2-\tilde r\int d\tau\cos \left(2\phi^a(\tau)\right)$ is the action of the boundary sine-Gordon model. Enforcing the constraint $\lambda = \sum_{a=1}^N\phi^a$ through a Lagrange multiplier $\rho$ yields $Z=\int \mathcal D \rho\, \mathcal{Z}[\rho]$ with
\begin{align}\label{eq:Z_rho}
	\mathcal{Z}[\rho]&=\int \mathcal D\bd \phi \mathcal D \lambda \exp\left[-\sum_{a=1}^{N}S_{\rm bsG}[\phi^a]  - \int_0^\beta d \tau E_C\left(\frac{\lambda(\tau)}{\pi} + \frac12\right)^2 + i \int_0^\beta d \tau \rho(\tau) \left(\lambda(\tau) - \sum_{a=1}^N\phi^a(\tau)\right)\right].
\end{align}
Observe that $\rho$ couples to $\sum_a \phi^a$ which is the sum of $N$ independent and identically distributed random variables. Owing to the central limit theorem, the fluctuations of $\sum_a \phi^a$ grow as $\sqrt{N}$ so performing a mean-field (or saddle-point) approximation for $\rho$ becomes exact in the limit $N\to \infty$. This consists in setting $Z=\mathcal{Z}[\rho^\star]$ with $\rho^\star$ such that $\frac{\delta \mathcal{Z}}{\delta \rho} [\rho^\star]=0$. Since $\mathcal{Z}[\rho]=\mathcal{Z}[-\rho]$ (to prove this, do the change of variables $\lambda \to -\pi - \lambda, \quad \phi^a \to -\phi^a - \pi \delta_{a,1}$ in Eq.~\eqref{eq:Z_rho}.), the mean-field solution is $\rho^\star=0$. Therefore, the partition function is
\begin{align}
	Z&\overset{N \gg 1}{=}\mathcal{Z}[\rho^\star]=\left[\int \mathcal D \phi\exp(-S_{\rm bsG}[\phi] )\right]^N \int \mathcal D \lambda\exp\left[- \int_0^\beta d \tau E_C\left(\frac{\lambda(\tau)}{\pi} + \frac12\right)^2 \right],
\end{align}
which is nothing but $N$ independent replicas of the boundary sine-Gordon model and a residual decoupled degree of freedom $\lambda$.

We also note that the mapping in the Toulouse limit $E_C \to \infty$ can be understood from a perturbative RG analysis of the QBM in the Toulouse limit $(E_c \to \infty)$ with asymmetric channels:
\begin{align}
	S=\frac{1}{\pi K}\sum_{\omega_n}|\omega_n||\bd \varphi(i\omega_n)|^2-\sum_{a=1}^{N}\tilde r_a\int_0^\beta d\tau\cos \left(2\bd G_a\cdot \bd \varphi(\tau)-\frac{\pi}{\sqrt{N}}\right).
\end{align}
The reflection amplitudes $r_{b(\ne a)}$ do not enter the connected correlation functions that appear in the $\beta$ function of $r_a$ up to order $N-1$. It is thus natural to believe that the system decouples into $N$ independent boundary sine-Gordon models in the limit where $N\to \infty$.

\setcounter{equation}{0}
\setcounter{figure}{0}
\setcounter{table}{0}
\renewcommand{\theequation}{S\arabic{equation}}
\renewcommand{\theHequation}{S\arabic{equation}}
\renewcommand{\thefigure}{S\arabic{figure}}	
\renewcommand{\bibnumfmt}[1]{[S#1]}
\renewcommand{\citenumfont}[1]{S#1}

\end{document}